\documentclass[debug]{rmaa}

\usepackage{hyperref}
\usepackage{rotating}
\usepackage{ulem}    

\usepackage{paralist}
\usepackage{tabularx,adjustbox,booktabs}

\usepackage{psfrag,color,ulem,gensymb}
\definecolor{deepblue}{rgb}{0,0,0.5}
\definecolor{deepred}{rgb}{0.6,0,0}
\definecolor{deepgreen}{rgb}{0,0.5,0}
\definecolor{darkgreen}{rgb}{0,0.6,0}

\usepackage[latin1]{inputenc}

\usepackage{textcomp}
\usepackage{xspace} 
\usepackage{booktabs,caption}
\usepackage[flushleft]{threeparttable}

\newcommand{\temp}{\hbox{T$_e$}} 

\newcommand{\tauv}{\hbox{$\tau_{V}$}}
\newcommand{\tauvi}{\hbox{$\tau_{V}^{o}$}}

\newcommand{\chin}{\hbox{$\chi^2_{\rm rno}$}}

\newcommand{\quof}{\hbox{\footnotesize{\textquotesingle\textquotesingle}}}

\newcommand{\lam}{\hbox{$\lambda$}}
\newcommand{\lamf}{\hbox{$\footnotesize{\lambda}$}}

\newcommand{\nout}{\hbox{\it n$_{H}^{\rm o}$}}

\newcommand{\nopa}{\hbox{\it n$_{\rm opa}$}}

\newcommand{\ncut}{\hbox{\it n$_{\rm cut}$}}
\newcommand{\nsng}{\hbox{\it n$_{\rm sng}$}}
\newcommand{\nlow}{\hbox{\it n$_{\rm low}$}}

\newcommand{\Tsii}{\hbox{\it T$_{\rm SII}$}}
\newcommand{\Toiii}{\hbox{\rm T$_{\rm OIII}$}}
\newcommand{\Tmean}{\hbox{\rm $\left<{\rm T}_{\rm OIII}\right>$}}
\newcommand{\Tosng}{\hbox{\rm T$_{\rm OIII}^{\,sng}$}}
\newcommand{\Tosngav}{\hbox{$\left<{\rm T}_{\rm OIII}^{\,sng}\right>$}}
\newcommand{\Tosnmn}{\hbox{$\left<{\rm T}_{\rm OIII}\right>$}}
\newcommand{\uout}{\hbox{\it {\rm U}$_{\rm o}$}}

\newcommand{\Tcut}{\hbox{\it T$_{cut}$}}

\newcommand{\cms}{\hbox{${\rm cm^{-2}}$}}

\newcommand{\epsi}{\hbox{$\epsilon$}}

\newcommand{\aox}{\hbox{$\alpha_{OX}$}}
\newcommand{\auv}{\hbox{$\alpha_{UV}$}}
\newcommand{\fauv}{\hbox{$\alpha_{FUV}$}}
\newcommand{\abunoh}{\hbox{$12 + \log({\rm O/H}$}}

\newcommand{\ztot}{\hbox{$Z_{\rm tot}$}}

\newcommand{\zsol}{\hbox{$Z_{\sun}$}}

\newcommand{\Vband}{\hbox{V-band}}

\newcommand{\degk}{{\degree}K}

\newcommand{\cmc}{cm$^{-3}$}
\newcommand{\kms}{km~s$^{-1}$}

\newcommand{\hii}{\hbox{H\,{\sc ii}}}

\newcommand{\hep}{\hbox{He$^{\rm +}$}}

\newcommand{\opp}{\hbox{O$^{\rm +2}$}}

\newcommand{\arppp}{\hbox{Ar$^{\rm +3}$}}

\newcommand{\neppp}{\hbox{Ne$^{\rm +3}$}}

\newcommand{\lya}{\hbox{Ly\,$\alpha$}}

\newcommand{\ha}{\hbox{H$\alpha$}}

\newcommand{\hb}{\hbox{H$\beta$}}

\newcommand{\hei}{\hbox{He\,{\sc i}}}
\newcommand{\heii}{\hbox{He\,{\sc ii}}}

\newcommand{\heiwr}{\hbox{He\,{\sc i}\,$\lambda$5876\AA}}

\newcommand{\heiwar}{\hbox{He\,{\sc i}\,$\lambda$4713\AA}}

\newcommand{\fblHe}{\hbox{$f^{HeI}_{\rm blend}$}}
\newcommand{\fblNe}{\hbox{$f^{NeIV}_{\rm blend}$}}
\newcommand{\Rhei}{\hbox{R$_{\rm He}$}}
\newcommand{\RHeAr}{\hbox{R$_{\rm He/Ar}^{}$}}
\newcommand{\RNeAr}{\hbox{R$_{\rm Ne/Ar}^{}$}}

\newcommand{\sii}{\hbox{[S\,{\sc ii}]}}
\newcommand{\siiw}{\hbox{[S\,{\sc ii}]\,\lam\lam6716,31\AA}}
\newcommand{\siirr}{\hbox{\lam6716\AA/\lam6731\AA}}
\newcommand{\siibr}{\hbox{\lam\lam4069,76\AA/\lam\lam6716,31\AA}}
\newcommand{\oiirr}{\hbox{\lam6726\AA/\lam6729\AA}}
\newcommand{\oiibr}{\hbox{\lam3727\AA/\lam7325\AA}}
\newcommand{\ariva}{\hbox{\lam4711\AA}}
\newcommand{\arivb}{\hbox{\lam4740\AA}}
\newcommand{\arivr}{\hbox{\lam4711\AA/\lam4740\AA}}

\newcommand{\arivtop}{\hbox{\lam4711\AA{\tiny +}/\lam4740\AA}}

\newcommand{\arivl}{\hbox{\lam\lam4711,40\AA}}

\newcommand{\neiva}{\hbox{\lam\lam4715\AA}}
\newcommand{\neivaa}{\hbox{\lam\lam4714,16\AA}}

\newcommand{\neivb}{\hbox{\lam\lam4725\AA}}
\newcommand{\neivbb}{\hbox{\lam\lam4724,26\AA}}

\newcommand{\oiiir}{\hbox{\lam4363\AA/\lam5007\AA}}

\newcommand{\niir}{\hbox{\lam5755\AA/\lam6583\AA}}

\newcommand{\heiirb}{\hbox{\lam4686\AA/\lam4861\AA}}

\newcommand{\ariv}{\hbox{[Ar\,{\sc iv}]}}
\newcommand{\arivp}{\hbox{[Ar\,{\sc iv}]{\tiny +}}}
\newcommand{\neiv}{\hbox{[Ne\,{\sc iv}]}}
\newcommand{\nev}{\hbox{[Ne\,{\sc v}]}}

\newcommand{\wavoo}{\hbox{(\lam4363\AA/\lam5007\AA)}}
\newcommand{\wavhb}{\hbox{(\lam5007\AA/\lam4861\AA)}}

\newcommand{\Rnii}{\hbox{$R_{\rm NII}$}}
\newcommand{\Roiii}{\hbox{$R_{\rm OIII}$}}

\newcommand{\oiv}{\hbox{[O\,{\sc iv}]}}
\newcommand{\oivw}{\hbox{[O\,{\sc iv}]\,\lam25.89\,$\mu$m}}

\newcommand{\oiii}{\hbox{[O\,{\sc iii}]}}
\newcommand{\oiiiw}{\hbox{[O\,{\sc iii}]\,\lam5007\AA}}
\newcommand{\oiiitw}{\hbox{[O\,{\sc iii}]\,\lam4363\AA}}

\newcommand{\Loiv}{L_{[O\,{IV}]}}
\newcommand{\oii}{\hbox{[O\,{\sc ii}]}}

\newcommand{\feii}{\hbox{Fe\,{\sc ii}}}

\newcommand{\fevii}{\hbox{Fe\,{\sc vii}}}

\newcommand{\feviiw}{\hbox{Fe\,{\sc vii}\,\lam6086\AA}}

\def\arcsec{\hbox{$^{\hbox{\rlap{\hbox{\lower4pt\hbox{$\,\prime\prime$}}
          }\hbox{$\frown$}}}$}}
\def\arcmin{\hbox{$^{\hbox{\rlap{\hbox{\lower4pt\hbox{$\;\prime$}}
          }\hbox{$\frown$}}}$}}

\newcommand{\citeNN}[1]{#1}   
\newcommand{\out}[1]{}  

\newcommand{\map}{\hbox{{\sc mappings i}g}}
\newcommand{\ifla}{\hbox{{\sc i}g}}
\newcommand{\MAP}{\hbox{{\sc mappings i}}}
\newcommand{\OSALD}{{\sc osald}}

\usepackage{lscape}
\usepackage{listings}
\DeclareFixedFont{\ttb}{T1}{txtt}{bx}{n}{10} 
\DeclareFixedFont{\ttm}{T1}{txtt}{m}{n}{10}  

\title{Temperature Discrepancy with Photoionization Models of the Narrow-Line Region}

\author{
  Luc Binette,\altaffilmark{1} 
  Montserrat Villar Mart\'in,\altaffilmark{2}
  Gladis Magris C.,\altaffilmark{3}
  Mariela Mart\'inez-Paredes,\altaffilmark{4}
  Alexandre Alarie,\altaffilmark{5}
  Alberto Rodr\'\i{}guez Ardila,\altaffilmark{6}
  and
  Ilhuiyolitzin Villica\~na-Pedraza\altaffilmark{7}
  }

\altaffiltext{1}{Instituto de Astronom\'\i{}a, UNAM, M\'exico}
\altaffiltext{2}{Centro de Astrobiolog\'\i{}a, Departamento de Astrof\'\i{}sica, Madrid, Spain}
\altaffiltext{3}{Centro de Investigaciones de Astronom\'\i{}a, M\'erida, Venezuela}
\altaffiltext{4}{Korea Astronomy and Space Science Institute, Daejeon, South Korea}
\altaffiltext{5}{D\'epartement de physique, de g\'enie physique et d'optique,  Universit\'e Laval, Qu\'ebec}
\altaffiltext{6}{Laborat\'orio Nacional de Astrof\'\i{}sica, Itajub\'a, Brazil}
\altaffiltext{7}{DACC Science Department, New Mexico State University, USA}

\shortauthor{Binette, Villar-Mart\'\i{}n, Magris et\,al.}

\shorttitle{Narrow-Line Region of Active Nuclei}

\fulladdresses{
\item L. Binette: Instituto de Astronom\'\i{}a, Universidad Nacional Aut\'onoma de M\'exico, A.P. 70-264, 04510 M\'exico, D.F., M\'exico,
Apartado Postal 70-264, Ciudad de M\'exico, CDMX, C.P. 04510, M\'exico. 
\item M. Villar-Mart\'\i{}n: Centro de Astrobiolog\'\i{}a, (CAB, CSIC-INTA), Departamento de Astrof\'\i{}sica, Cra. de Ajalvir Km. 4, 28850, Torrej\'on de Ardoz, Madrid, Spain.
\item G. Magris: Centro de Investigaciones de Astronom\'\i{}a, Apartado Postal 264, M\'erida 5101-A, Venezuela
\item M. Mart\'inez-Paredes: Korea Astronomy and Space Science Institute 776 Daedeokdae-ro, Yuseong-gu, Daejeon, 34055, Republic of Korea.
\item A. Alarie: D\'epartement de physique, de g\'enie physique et d'optique,  Universit\'e Laval, Qu\'ebec, QC G1V 0A6, Canada.
\item A. Rodriguez Ardila: Laborat\'orio Nacional de Astrof\'\i{}sica - Rua dos Estados Unidos 154, Bairro das Na\c{c}\~oes. CEP 37504-364, Itajub\'a, MG, Brazil.
\item I. Villica\~na-Pedraza: DACC Science Department, New Mexico State University, Las Cruces, NM 88003, USA.
  }

\listofauthors{L. Binette, I., G. Magris, M. Villar-Mart\'in, M. et\,al.}
\indexauthor{Binette, L.}
\indexauthor{Magris, G.}
\indexauthor{Villar-Mart\'in, M.}
\indexauthor{Mart\'inez-Paredes , M.}
\indexauthor{Alarie, A.}
\indexauthor{Rodr\'iguez Ardila, A.}
\indexauthor{Villica\~na-Pedraza, I.}
\abstract{Using published work on the narrow-line region of Active Galactic Nuclei, a comparison is carried out among the \oiii\  \oiiir\ (\Roiii) {ratio} observed in quasars, Seyfert\,2's and the spatially resolved ENLR plasma. Using the  weak \ariv\ \arivr\ doublet {ratio} observed by  Koski (1978) among Seyfert\,2's, we find evidence of a  Narrow-Line Region (NLR) populated by low density emission clouds  ($\la 10^4\,$\cmc). After considering calculations of the \ariv\ and \oiii\ ratios that assume a powerlaw distribution of plasma densities, no evidence of collisional deexcitation is found. The plasma temperature that is inferred is 13\,500\,\degk, which is problematic to reproduce with standard photoionization calculations. The simplest interpretation for the near coincidence of the \Roiii\ ratios among the ENLR and Seyfert\,2  measurements ($\Roiii \simeq 0.017$) is that the low density regime applies to both plasmas. 
}

\resumen{Utilizando trabajos publicados sobre la regi\'on de l\'{\i}neas angostas de los N\'ucleos Gal\'acticos Activos, se realiza una comparaci\'on entre el \oiii\  \oiiir\ (\Roiii) observado en cu\'asares, Seyfert\,2 y en el plasma espacialmente resuelto de la ENLR. Utilizando el d\'ebil doblete de \ariv\ \arivr\ observado por Koski encontramos evidencias de una regi\'on de l\'{\i}neas angostas (NLR) poblada por nubes de emisi\'on de baja densidad ($\la 10^4\,$\cmc). Tras considerar los c\'alculos de las relaciones \ariv\ y \oiii\ que asumen una distribuci\'on de ley de potencia de las densidades del plasma, no se encuentra evidencia de deexcitaci\'on colisional. La temperatura del plasma que se infiere es de 13\,500\,\degk, la cu\'al es dificil de reproducir con los c\'alculos est\'andar de fotoionizaci\'on. La interpretaci\'on m\'as sencilla de la casi coincidencia de los cocientes de \Roiii\ medidos en la  ENLR y las Seyfert\,2 ($\Roiii \simeq 0.017$)  ser\'{\i}a que el r\'egimen de baja densidad se aplica a ambos plasmas.
}

\addkeyword{quasars: emission lines}
\addkeyword{galaxies: Seyfert}
\addkeyword{dust: extinction}
\addkeyword{plasmas}

\begin{document}
\maketitle


\section{Introduction}
\label{sec:intro}

The physics of the so-called narrow-line region of Active Galactic Nuclei (AGN) has been amply studied  \citep[cf.][and references therein]{OST78}. AGN emission line spectra can be divided into two categories\footnote[1]{The current study does \textit{not} include BL\,Lac objects nor Extremely Red Quasars.}: Type\,I when the full width half-maximum (FWHM) of the permitted lines are significantly larger than the forbidden lines, and Type\,II where both the permitted and forbidden lines have similar FWHM. The broad-line region (BLR) observed in Type\,I objects originate from high density gas ($>10^8$\,\cmc) much closer to the black-hole (BH) than the narrow-line region (hereafter NLR), the latter being observed in both Type\,I and II objects. The AGN unified model sustains that both types relate to the same phenomenon, with the differences being the visibility of the central engine. It proposes that the BLR is hidden from direct view in Type\,II due to an optically thick dusty torus-like gas structure surrounding the central engine (black hole, accretion disk, and BLR) \citep{An93}. Whether the BLR is observed or not depends on the viewing angle of the nucleus. Seyfert\,2's, QSO\,2's and Narrow Line Radio Galaxies (NLRG) are classified as Type\,II while Seyfert\,1's, quasars, QSO\,1's and Broad-Line Radio Galaxies (BLRG) are of Type\,I because their BLR is visible. 

While it is customary for \hii\ regions to assume the low density regime (hereafter LDR) when evaluating the plasma temperature using the \oiii\ \oiiir\ line ratio (hereafter labeled \Roiii), this is inappropriate with the NLR, at least in Type\,I objects. \citet{OST78} interpreted the relative strength of the $\lambda 4363$\AA\ line, which was measured to be higher in Seyfert\,1's and BLRG than in Seyfert\,2's, as evidence of densities in the range $10^6-10^7$\,\cmc\ within the NLR of Type\,I AGN. This interpretation was confirmed by the study of \citet[hereafter BL05]{BL05} who compared the \Roiii\ they measured in 30 quasars. 
Their single-density calculations showed that the broad range of observed \Roiii\ ratios implies high plasma densities ranging from possibly $10^5$ up to $10^7$\,\cmc, providing convincing evidence of the important role of collisional deexcitation in Type\,I AGN where the temperature cannot be directly inferred from the \Roiii\ ratio.  In the case of Type\,II objects (Seyfert\,2's and NLRG), the \Roiii\ ratio is on average smaller ($\la 0.019$)  although selection effects may possibly bias such assessment. Prevailing NLR photoionization models consider a distribution of clouds that extends over a wide range of densities and ionization parameter values, whether the targets are Type\,I \citep{Ba95,Kor97a} or Type\,II objects \citep{Fg97,Ri14}. With respect to the spatially resolved emission line component of AGN, the so-called extended NLR (hereafter ENLR), it consists of \textit{off-nuclear} line emission  from plasma at typically LDR densities \citep[e.g.][]{Ta94,Be06a,Be06b} where the \Roiii\ ratio should provide a reliable temperature measurement.

The original element of the current work is the use of the weak \ariv\ $\lambda\lambda$4711,40\AA\ doublet to evaluate to what extent the \Roiii\ measurements of our selected Seyfert\,2 sample is affected by collisional deexcitation.  To cover the multi-density case, we developed an algorithm, \OSALD, to calculate density and temperature line ratio \textit{diagnostics} appropriate to isothermal plasmas in which the density follows a powerlaw distribution rather than taking on a single value.  This algorithm offers the option of including a foreground dust extinction component whose opacity, rather than being uniform, correlates with the emission plasma density. Our main conclusion is that, at least for the subset of Type\,II objects where the \ariv\ doublet is observed, there is no evidence of significant collisional deexcitation. Hence in those cases the \Roiii\ ratio constitutes a direct temperature indicator. Oddly, LDR photoionization calculations result in temperature discrepancies with the observations, underscoring the so-called temperature problem \citep{SB96,Be06a,VM08,Dr15,Dr20b}. In a follow up Paper, we evaluate different physical processes to address this issue.


\begin{figure}[!t]
\includegraphics[width=\columnwidth]{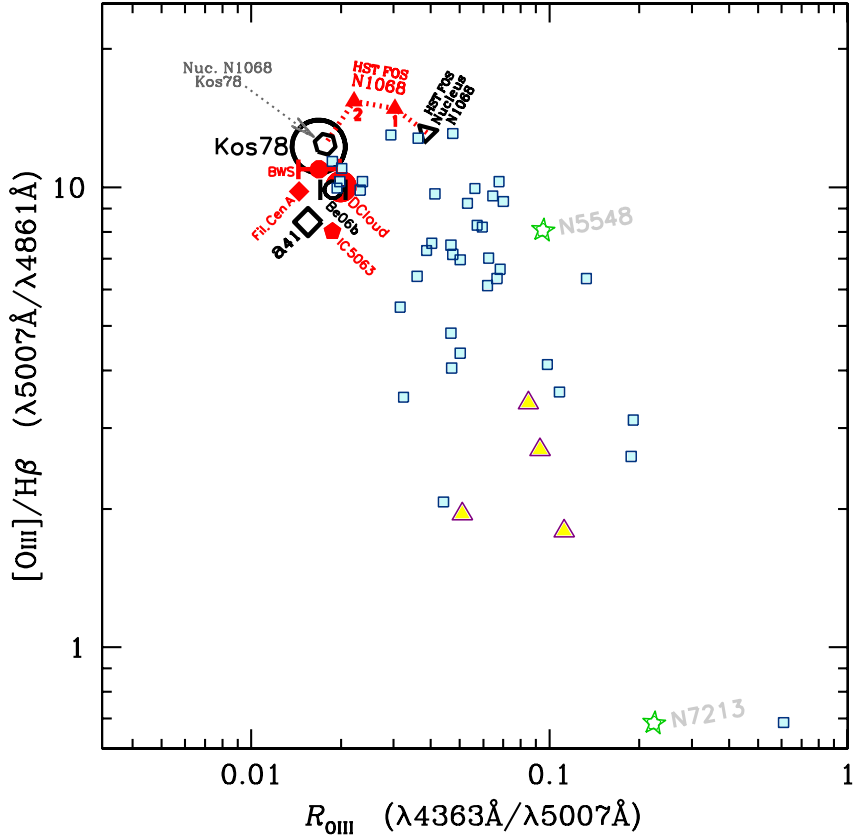}
\caption{AGN dereddened line ratios of \oiii/\hb\  
vs. \Roiii\ from: 
A- Type\,I AGN with measurements of 1) 30 quasars studied by \citeNN{BL05} (bluish open squares), 2) four narrow-line Seyfert\,1 galaxies from RA00 (yellowish open triangles) 3) two Seyfert\,1.5, NGC\,5548 and NGC\,7213 (open stars), B-Type\,II AGN represented by open \textit{black} symbols consisting of 1) the average of seven Seyfert\,2's from Kos78 (large circle), 2) the average of four Seyfert\,2's from Be06b (small circle), 3) the high excitation Seyfert\,2 subset a41 from \citeNN{Ri14} (diamond), 4) the nucleus of NGC\,1068 through ground-based observations by Kos78 (black hexagon) and HST-FOS observations analysed by Kr98 (black triangle), and C- ENLR measurements (all with red filled symbols) consisting of 1) the average from BWS of two Seyfert\,2's and two NLRGs (red dot), 2) the long-slit observations of the Seyfert\,2 IC\,5063 by Be06b (pentagon), 3)  the average of seven spatially resolved optical filaments from the radio-galaxy Centaurus\,A  (red square) by Mo91, 4) the 8\,kpc {\it distant cloud} from radiogalaxy Pks\,2152$-$699 by Ta87 (large dot), and 5) the HST-FOS measurements of two ENLR knots from NGC\,1068 (red triangles).
}
\label{fig:fig1}
\end{figure}

Our reference sample is described in \S\,\ref{sec:obs} and a comparison with single-component photoionization models is presented in \S\,\ref{sec:model}. A modified interpretation of the NLR \Roiii\ ratios observed near $\simeq 0.017$ in Type\,II and some Type\,I AGN is proposed in \S\,\ref{sec:interp}. These are subsequently compared with calculations made with the algorithm \OSALD\ (\S\,\ref{sec:isot}), which considers a powerlaw density distribution.


\section{Reference data set of \Roiii\  ratios in AGN}
\label{sec:obs} 

In what follows, the term NLR will be used exclusively in reference to the spatially {\it unresolved} nuclear component. For any line emission that originates beyond the spatially unresolved central component\footnote{Where the densest plasma of the BLR and inner NLR is located}. of the active nucleus, it will be referred as ENLR\footnote{When the line emission lies kiloparsecs or more away from the nucleus, some authors \citep[e.g.,][]{TFSA} prefer the term Extended Emission Line Region (EELR).} in all cases where the gas is deemed photoionized by the AGN rather than by hot stars.

In order to evaluate the impact of collisional-deexcitation on the \oiii\ emission lines among Type\,I and II AGN, our data set consists primarily of the quasar sample of BL05 (excluding upper limits data), to which we added the four narrow-line Seyfert\,1 studied by \citet[hereafter RA00]{RA00} which were  originally observed by \citet{RAPD}. To have access to measurements of the $\lambda\lambda$4711,40\AA\  doublet, we relied on the Seyfert\,2 sample of \citet[hereafter Kos78]{Kos78}. Finally, to ensure that our sample covers cases where the emission plasma is negligibly affected by  collisional deexcitation, we included diverse ENLR observations from the literature. 
Fig.\,\ref{fig:fig1} describes the behaviour of the dereddened \oiii/\hb\ \wavhb\ and \Roiii\ \wavoo\ line ratios of our AGN sample.

\subsection{Detailed description of the dereddened \Roiii\  data set}
\label{sec:data} 

The data set was extracted from the following sources:
 \renewcommand{\labelenumi}{\Alph{enumi}}
 \begin{enumerate}

   \item -- NLR of Type\,I AGN
   \begin{enumerate}

     \item Based on the prominent work of BL05, the sample consists of 30 Type\,I quasars with $z < 0.5$, mostly from the bright quasar survey of \citet{BG82}. Objects where only upper limits of \oiiitw\ were reported have been excluded. The sample is represented by bluish open squares in Fig.\,\ref{fig:fig1}. The authors used the \oiiiw\ profile of each object as template for extracting the NLR \hb\ and \oiiitw\ line fluxes. Since the latter line is weak, its measurement required a proper subtraction of the underlying \feii\ emission multiplets. \citeNN{BL05} used the I\,Zw\,1 \feii\ template provided by T. Boroson (private communication) to subtract the \feii\ multiplets.  All the line fluxes were corrected for dust reddening and possible slit losses. 

     \item The measurements of the four narrow-line Seyfert\,1 nuclei (hereafter NLS1) from RA00 (yellowish open triangles) were annexed. As detailed in their study, the authors used their own spectrum of I\,Zw\,1 to subtract the various \feii\ underlying features present in their NLS1 spectra. They compared different ways to extract the \hb\ NLR contribution, favoring in the end the procedure of fitting a narrow and broad gaussian component to the \hb\ profiles. The broad to narrow \hb\ flux ratios in these objects cover the range of 1.8 to 3.4. 

     \item For comparison purposes, we included the measurements of two well-studied
      Seyfert\,1.5 galaxies (light-green open stars): NGC\,5548 \citep{KCFP} and  NGC\,7213\footnote{Initially associated to the LINER category, the presence of \nev\ lines indicates a high ionization plasma despite having \oiii/\hb\ $< 1$ due to collisional deexcitation. }  \citep{FH84}.

   \end{enumerate}

   \item -- NLR of Type\,II AGN 
   \begin{enumerate}

     \item  To characterize the behaviour of high excitation Type\,II objects, we adopt the pioneering work on Seyfert\,2's by \citet[hereafter Kos78]{Kos78}, which provides the unique characteristic of reporting reliable measurements of the weak \ariv\ $\lambda\lambda$4711,40\AA\ doublet ratio, an essential density indicator for evaluating in \S\,\ref{sec:sng} and \S\,\ref{sec:fit} to what extent the observed \Roiii\ is affected by collisional deexcitation. 
     Table\,\ref{tab:sam} lists the reddening corrected ratios of the high excitation subset of their sample (i.e. with $\oiii/\hb\ \ge 10$), which consists of seven Seyfert\,2's. Two objects, Mrk\,348 and 3C33, were left out of Table\,\ref{tab:sam} since their measurement of the \ariv\  ratio unrealistically exceeded the low density limit value. They presumably indicate emission from LDR plasma. The average \Roiii\ ratio from Table\,\ref{tab:sam} is 0.0168 (i.e. $10^{-1.77}$ in Fig.\,1), which is represented by a large black disk whose radius of 0.088\,dex corresponds to the \Roiii\ RMS dispersion. The average \oiii/\hb\ is $12.3 \pm 1.1$. 

     \item As a complement to Type\,II objects, we averaged the measurements of the four Seyfert\,2's IC\,5063, NGC\,7212, NGC\,3281 and NGC\,1386 observed by \citet[][hereafter Be06b]{Be06b}. It is represented by  a small black circle corresponding to a mean \Roiii\ of 0.0188. Pseudo error bars represent the RMS dispersion of 0.042\,dex.
     
    \item The black diamond labelled a41 with $\Roiii=0.0155$ represents the high ionization end of the sequence of reconstructed spectra of \citet[][hereafter Ri14]{Ri14} which was extracted from a sample of 379 AGN. These were identified by applying the Mean Field Independent Component Analysis (MFICA) tool to their Sloan Digital Sky Survey (SDSS) sample of $\sim 10^4$ emission line galaxies in the redshift range $0.10 < z < 0.12$ \citep[see also][]{Allen2013}. They meticulously reviewed each spectrum to ensure that no BLR component was present.
    \item Ground-based observations of the Seyfert\,2 NGC\,1068 nucleus by Kos78 is represented by the black open octagon while the black open triangle corresponds to the HST-FOS measurement of the nucleus at a much higher spatial resolution of 0.3\arcsec\ \citep[archive data][hereafter Kr98]{Kr98}. 

   \end{enumerate}

   \item -- Spatially resolved ENLR emission
   \begin{enumerate}

     \item The red filled dot stands for the average ratio from the ENLR of four Type\,II AGN (two are Seyfert\,2's: ESO\,362-G08 and MRK\,573, and two are NLRGs: Pks\,0349$-$27 and Pks\,0634$-$20) which were studied by \citet[hereafter SB96]{SB96}. The mean \Roiii\ ratio is $0.0169 \pm 0.0029$,  which includes measurements on both sides of the nucleus, except for ESO\,362-G08 \citep[hereafter BWS]{BWS}. Pseudo-error bars denote an RMS dispersion of 0.07\,dex. 

     \item The Seyfert\,2 IC\,5063. The red pentagon represents the average ratio $\Roiii=0.0188$ (with dispersion of 0.042\,dex) from the {\it extranuclear} radial emission of the  Seyfert\,2 IC\,5063 which Be06b observed with a $S/N > 3$ from 8\arcsec\ NW to 5\arcsec\ SE.
     
     \item The Centaurus\,A (NGC\,5128) filaments. The red square represents the average ratio of seven optical filaments studied by \citet[hereafter Mo91]{Mo91} and situated along the radio jet at a mean distance of 490\,pc from the nucleus of the radio-galaxy Centaurus\,A (mean $\Roiii=0.0145$ with a dispersion of 0.13\,dex).

     \item Detached cloud emission aligned with radio-galaxy jets. The large red dot represents the well studied 8\,kpc distant cloud associated to the nucleus of radiogalaxy Pks\,2152$-$699 \citep[hereafter Ta87]{Ta87}.

    \item Two detached emission line 'knots' of NGC\,1068 labelled 1 and 2 (red triangles) which were studied by Kr98 using HST-FOS archive data. The positions are off-centered from the nucleus by 0.2\arcsec\ and 0.7\arcsec, respectively. 

  \end{enumerate}
 \end{enumerate}


\section{Photoionization calculations at LDR densities}
\label{sec:model} 

All our calculations will be presented in Fig.\,\ref{fig:fig3} in which the reference data set is represented using the same symbols but colored in gray. 
For an isothermal plasma at a fixed temperature, densities much above LDR would cause an increase of the \Roiii\ ratio, shifting its position to the right in Fig.\,\ref{fig:fig3} due to collisional deexcitation.  The segmented cyan arrow describes the increase in \Roiii\ expected from a 14\,000\,\degk\ plasma slab whose density successively takes on the values of $10^2$ (LDR), $10^{4.5}$, $10^5$, $10^{5.5}$, $10^{6}$ and $10^{6.5}$\,\cmc. It illustrates the density range implied by the BL05 and RA00 Type\,I AGN if they shared the same temperature. For a 15\,000\,\degk\ plasma, the critical densities\footnote{We define the critical density as the density where the line intensity, divided by both the ion and the electron densities, reaches 50\% of the low density limit value.} for deexcitation of the \oiii\ \lam5007\AA\ and \lam4363\AA\ lines are $7.8\times 10^5$ and $2.9\times 10^7$\,\cmc, respectively, while for the\ariv\ \lam4711\AA\ and \lam4740\AA\ lines these are $1.7 \times 10^4$ and $1.5\times 10^{5}$\,\cmc. Let us compare the data with the values predicted by photoionization calculations. 


\begin{figure}[!t]
  \includegraphics[width=\columnwidth]{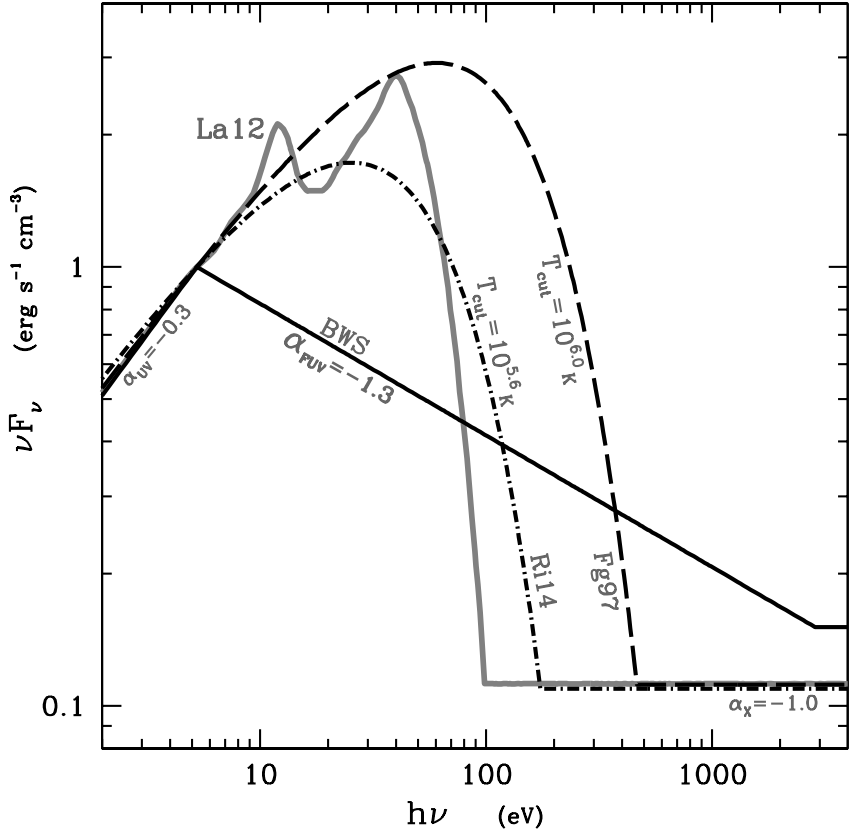}
  \caption{The four SEDs  described in  \S\,\ref{sec:model} and adopted in our LDR single-component photoionization models of Fig.\,\ref{fig:fig3}: 1) the SED assumed by Ri14 with $\Tcut = 10^{5.62}$\,\degk\ in their LOC model calculations (dot-dashed line), 2) a similar SED but with a higher \Tcut\ of $10^{6.0}$\,\degk\ as explored by Fg97 (long dashed-line), 3) a powerlaw SED with $\fauv=-1.3$ adopted by BWS (black continuous line), 4) the double bump thermal SED proposed by La12 (thick gray line). Each SED is expressed in $\nu F_{\nu}$ units and normalized to unity at 5\,eV (2000\AA).  In the X-rays, they all convert into a powerlaw of index $-1.0$. 
  }
\label{fig:fig2}
\end{figure}

\subsection{Above solar gas metallicities}
\label{sec:metal}

The abundances we adopt correspond to 2.5\,\zsol, a value within the range appropriate to galactic nuclei of spiral galaxies. For instance, the landmark study by \citet{Do14} of multiple \hii\ regions of the Seyfert\,2  NGC\,5427 favor abundances significantly above solar. Using the Wide Field Spectrograph \citep[WiFeS:][]{Do10}, the authors could determine the ISM oxygen radial abundances using 38 \hii\ regions spread between 2 and 13\,kpc from the nucleus. Using their inferred metallicities, they subsequently modelled the line ratios of over 100 `composite' ENLR-\hii\ region emission line spaxels as well as the line ratios from the central NLR. Their highest  oxygen abundance reaches 3\,\zsol\  (i.e. $\abunoh = 9.16$). Such a high value is shared by other observational and theoretical studies that confirm the high metallicities of Seyfert nuclei \citep{SP90,Na02,Ba08}. \textbf{Our abundance set relative to H corresponds to twice the solar values} of \citet{AG06} except for C/H and N/H, which are set at four times the solar values owing to secondary enrichment, resulting in a gas metallicity of $\ztot =2.47 \zsol$ by mass. For the He/H abundance ratio, we assume the solar value of 0.085.

\subsection{Four alternative ionizing energy distributions}
\label{sec:sed}

The four spectral energy distributions (hereafter SED)  selected are shown in Fig.\,\ref{fig:fig2}. They are representative of published work concerning AGN photoionization models and can be described as follows:
\renewcommand{\labelenumi}{\Alph{enumi}}
 \begin{enumerate}

\item the long dashed-line represents the SED used by \citet[hereafter Fg97]{Fg97} in their calculations of  Local Optimally emitting Cloud (LOC) models for the NLR. It is characterized by a thermal bump of the form $$F_{\nu} \propto \nu^{\auv} \exp(-h\nu/k\Tcut)$$ with $\auv=-0.3$ and $\Tcut = 10^{6.0}$\,\degk. 

\item  the dot-dashed line represents the 'optimized SED' used by \citet[hereafter Ri14]{Ri14} in their calculations of LOC models. It shares the same index \auv\ as the Fg97 SED above  but assumes a lower \Tcut\ of $10^{5.52}$\,\degk\ to describe the thermal bump. It is significantly softer than the Fg97 SED, yet due to collisional deexcitation being important in LOC models, these qualitatively reproduces the \Roiii\ and \heii/\hb\ (\heiirb) ratios of their reconstructed Seyfert\,2 emission line spectra.

\item a powerlaw of index $\fauv=-1.3$ in the far-UV domain (continuous black line) which was used by \citet[hereafter BWS]{BWS} and \citet{BWRS} in their matter-bounded cloud calculations.

\item the thick continuous gray line represents the sum of two distinct thermal bumps as proposed by \citet[hereafter La12]{La12} who postulated that the accretions is entirely covered by intervening thick BLR clouds, which would absorb \lya\ as well as the softer ionizing radiation, thereby accounting for the observed UV steepening short-ward of 1050\AA. The author proposed that the absorbed EUV energy is reprocessed into emission at much shorter wavelengths, which generates the second peak near 40\,eV.
\end{enumerate}

Each SED converts in the X-rays into a powerlaw $F_{\nu} \propto \nu^{-1}$ and results in an \aox\ index\footnote{Defined by the flux ratio at 2\,keV with respect to 2500\AA.} of $-1.35$ except  the BWS SED ($\aox\ = -1.30$).


\begin{figure}[!t]
\includegraphics[width=\columnwidth]{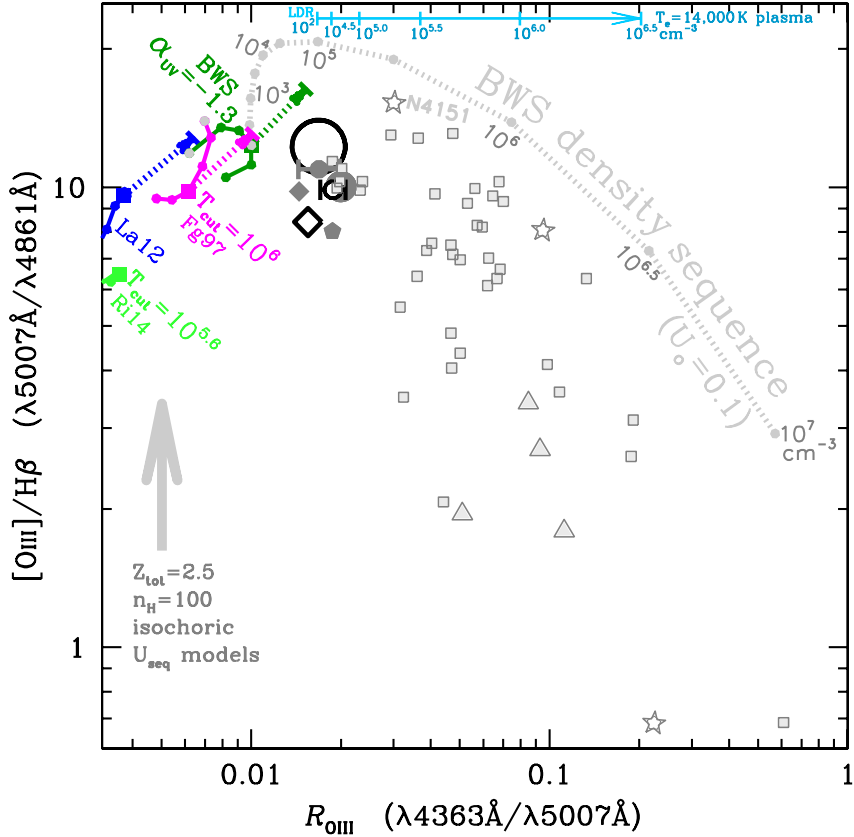}
\caption{Same data set from Fig\,\ref{fig:fig1} except that it now includes the Seyfert\,1 NGC\,4151 (c.f. \S\,\ref{sec:evaneiv}) instead of NGC\,1068. The gray open symbols correspond to the NLR from Type\,I objects, black symbols to Type\,II and gray filled symbols to ENLR measurements. Four sequences of LDR photoionization models are shown (solid lines) along which the ionization parameter \uout\ increases in steps of 0.5\,dex, from 0.01 (light gray dot) up to 0.46. Some models fall outside the Figure limits. The sequences are labelled according to the SED defined in \S\,\ref{sec:sed}: A) Fg97 (magenta), B) Ri14 (light green), C) BWS (dark green) and D) La12 (blue). The models are all ionization bounded, dustfree, isochoric with $\nout = 100$\,\cmc and  2.5\,\zsol\ abundances. A square identifies the $\uout = 0.1$ model with a dotted arrow representing the shift when one adopts the 1.5\,\zsol\ abundances of Ri14. The gray dotted line represents a density sequence in which the densities of the BWS model with $\uout=0.1$ are successively increased in steps of 0.5\,dex. The cyan segmented arrow at the top illustrates the effect of collisional deexcitation on the \Roiii\ ratio from a 
14\,000\,\degk\ plasma at successively larger densities. }
\label{fig:fig3}
\end{figure}

\subsection{Isochoric single-component photoionization calculations}
\label{sec:phot}

Sequences of photoionization models are shown in Fig.\,\ref{fig:fig3} corresponding to the four SEDs of Fig.\,\ref{fig:fig2}. LDR was assumed as it is the appropriate density regime for the ENLR plasma and for at least a significant subset of the AGN sample, as argued in \S\,\ref{sec:interp}. They were calculated using the most recent \footnote{This version includes the new algorithm \OSALD\ (App.\,\ref{sec:ap-o-subC}). Other updates are described in a subsequent paper \citep{Bi22b}. } version \ifla\ of the code \MAP\ \citep{BM12}  All models share the same density of $\nout=100$\,\cmc\ and each sequence includes up to six models\footnote{Some of the leftmost models of La12 and Ri14 fall off the graph boundaries.} along which the ionization parameter \uout\ increases in steps of 0.33\,dex, from 0.01 (gray dot) up to 0.46. A filled square denotes the $\uout =0.1$ model. All calculations are ionization-bounded, dustfree and isochoric. The observational data set represented in Fig.\,\ref{fig:fig3} uses only black or gray tones but with the same symbol coding as in Fig.\,\ref{fig:fig1}.  

If we define the photoheating efficiency of each SED as the temperature of the plasma averaged over the region occupied by the \opp\ ion, we obtain for the $\uout=0.1$ model the following values of 11\,300, 9700, 8400 and 8320\,\degk, assuming the SED which we labelled BWS, Fg97, La12 and Ri14, respectively (Fig.\,\ref{fig:fig3}). The BWS SED possess the highest efficiency but is more speculative as it excludes the possibility of thermal dump (or peak) in the far-UV. Such a feature is to be expected if the continuum originates from an accretion disk, which is widely accepted as being the primary mode of energy generation in AGN. The double-peak reprocessed SED of La12 presents the advantage of accounting for the `universal' knee observed at 10\,eV. The position of the second at 40\,eV, however, would need to be shifted to higher energies in order to increase the photoheating efficiency.

It is currently not possible to determine which abundances are the most appropriate to the environment of active nuclei although it is generally accepted that metallicities above solar are most likely. If one assume the more conservative metallicity of 1.5\,\zsol\ of Fg97 and Ri14, a shift towards the right takes place, as shown by the dotted line arrows in Fig.\,\ref{fig:fig3}, assuming an ionization parameter of $\uout=0.1$. 

What is the origin of the gap between the photoionization models and  AGN observations?  The positions of models on the left of Fig.\,\ref{fig:fig3} correspond to LDR conditions. It is a justified option for the ENLR, as argued in Appendix\,\ref{sec:ap-enlr}. As shown by BL05 as well as by the density sequence using the BWS SED in Fig.\,\ref{fig:fig3} (gray dotted line), all models can be shifted towards higher \Roiii\ values by assuming plasma densities much above $10^{4.3}\,$\cmc. This is the main reason why direct measurements of the density governing the \oiii\ lines are so important if we wish to determine the NLR temperature.  Measurements of the weak \ariv\ doublet can give us access to this information as explored below.

\section{Might the AGN buildup near $\Roiii \simeq 0.018$ represent a floor temperature?}
\label{sec:interp} 

The work of BL05 presented convincing evidence that the quasars (open squares) with \Roiii\ reaching $\sim 0.2$  is the manifestation of collisional deexcitation from high density plasma. Their single-density photoionization calculations suggest densities of $\simeq 10^{6.5}$\,\cmc. Interestingly, the four quasars on the extreme left appear to clump at $\Roiii = 0.0195$ with $\oiii/\hb\ \simeq 10.5$ in Fig.\,\ref{fig:fig3}. At this position, single-density photoionization calculations from BL05 suggest densities near $10^{5.2}$\,\cmc. If we consider an isothermal 14\,000\,\degk\ plasma (c.f. cyan arrow), the density we infer is very close to LDR, at $10^{3.3}$\,\cmc.
Our initial analysis of \Roiii\ among Seyfert\,2 galaxies indicated that these show \Roiii\ values similar to the leftmost quasars, which made us question whether collisional deexcitation is related to their position in Fig\,\ref{fig:fig3}. Our analysis of the NLR data of Seyfert\,2 galaxies, however, lead us to question whether collisional deexcitation is related to the position of these quasars on the left. It is noteworthy that For instance, a similar position is occupied by: 1) the sample of four Seyfert\,2 nuclei of Be06b (dark-green dot), 2)  the reconstructed Seyfert\,2 subset a41 from Ri14 (black diamond) which is based on an ample sample of SDSS spectra, and 3) the seven high excitation Seyfert\,2's of Kos78, as shown by the black disk, which represents the average \Roiii.
We would argue that the accumulation of AGN on the left is most likely representing a floor AGN temperature where collisional deexcitation is not significant.  To support our hypothesis, we will make use in \S\,\ref{sec:sng} of the density indicator provided by the \arivr\ doublet ratios of Kos78. 

\subsection{Why do ENLR observations coincide with the leftmost NLR \Roiii\ observations? }
\label{sec:why}

Because ENLR emission operates in the low density regime (c.f. App.\,\ref{sec:ap-enlr}), they provide ideal measurements to compare Type\,II objects with. In Fig.\,\ref{fig:fig1}, we added the following spatially resolved ENLR emission measurements:   1) the Cen\,A filaments (red square, Mo91), 2) the average ENLR ratios of four Type\,II AGN (red dot, BWS),  3) the radial ENLR emission from the Seyfert\,2 IC\,5063 (red pentagon: Be06b), 
and 4) the ionized cloud 8\,kpc distant from of Pks\,2152-69  \citep[large red dot,][]{Ta87}.  All gather rather relatively close to the leftmost Type\,I quasars (open squares) as well as to the mean Seyfert\,2 ratio of the Kos78 study (large black circle). 
Interestingly, the Seyfert\,2 reconstructed subset a41 (black diamond) of Ri14 occupies a similar position. The simplest interpretation would be that collisional deexcitation is \textit{not} significant, not only within the spatially resolved ENLR but among the leftmost objects as well, and that they all share a similar electronic temperature.


\begin{table}
\caption{Reddening-corrected Seyfert\,2 ratios from Koski (1978)}
\label{tab:sam}
\begin{changemargin}{-0.5cm}{-0.5cm}
\centering
\small
\begin{tabular}{clccccccc}
\toprule
\hspace{0.015cm} (1) & \hspace{0.225cm} (2)\tabnotemark{a} & (3) & (4) & (5) & (6) & (7)\tabnotemark{b} & (8)\tabnotemark{c} & (9)\tabnotemark{d} \\
Index & Seyfert\,2 & \oiii/H$\beta$  & \Roiii & \ariv{\tiny +} & \RHeAr & \fblHe  & \nsng & \Tosng \\ 

 \# &   & $\frac{\lamf5007}{\lamf4861}$  & $\frac{\lamf4363}{\lamf5007}$  & $\frac{\lamf4711{\tiny +}}{\lamf4740}$ & $\frac{\lamf5876}{\lamf4740}$ &  &  \cmc  & \degk  \\ 
\cmidrule{1-9}

1 & Mrk\,573  & 12.12 & 0.0149 & 1.167 & 1.52  & 0.039 & $1.85\times 10^3$ & 13\,360 \\ 

2 & Mrk\,34  & 11.46  & 0.0131 & 1.203 & 2.03 & 0.051 & $1.64\times 10^3$ & 12\,720 \\ 

3 & Mrk\,78  & 11.94 & 0.0117 & 1.267 & 2.22 & 0.053 & $1.14\times 10^3$ & 12\,210 \\ 

4 & Mrk\,176  & 14.36 & 0.0223 & 1.045 & 0.45 & 0.013 & $2.84\times 10^3$ & 15\,940 \\ 

5 & Mrk\,3   & 12.67 & 0.0189 & 0.837 & 1.95 &  0.072 & $7.21\times 10^3$ & 14\,670 \\ 

6 & Mrk\,1   & 10.95 & 0.0192 & 0.825 & 1.59 & 0.059 & $7.25\times 10^3$ & 14\,760 \\ 

7 & NGC\,1068  & 12.42 & 0.0177 & 0.790 & 2.42 & 0.097 & $8.77\times 10^3$ & 14\,210 \\ 


      \bottomrule
    \end{tabular}
  \end{changemargin}
    \begin{tablenotes}
      \small
      \item $^a$ The line ratios from the 7 Seyfert\,2's were reddening corrected by \citet{Kos78} using the observed Balmer decrement. The measurement uncertainties were estimated at $\pm 10\%$ for the strong line fluxes and $\pm 20\%$ for the weak lines. 
      \item $^b$ The inferred  fractional contribution of \heiwar\ to the blended \lam4711\AA{\tiny +} line.
      \item $^c$ The densities \nsng\ were determined using the deblended \arivr\ doublet ratio. 
      \item $^d$ The temperature \Tosng\ was derived from the \Roiii\ ratio assuming the density \nsng\ inferred from the deblended \ariv\ ratio (see \S\,\ref{sec:argon}). 
      The average temperature from Col.\,(7) is $\Tosngav =13\,980\pm 1200$\,\degk. 
    \end{tablenotes}
\end{table}

\subsection{Combination of the \ariv\ and \oiii\ diagnostics }
\label{sec:sng}

To evaluate the NLR density, we will rely on the observations of Kos78 who measured the weak \ariv\ $\lambda\lambda$4711,40\AA\ doublet of his Seyfert\,2 sample, an unusual feature among AGN surveys. The observations were carried out with the image-dissector-scanner mounted on the 3\,m Shane telescope at the Lick Observatory. The integration times were typically 32\,min (A. T. Koski, PhD thesis 1976). In Cols.\,(3--4) of Table\,\ref{tab:sam}, we present the reddening-corrected line ratios of \oiii/\hb\ and  \Roiii. In the context of high excitation planetary nebulae. \citet[][]{Ke19} pointed out that the \ariv\ and \oiii\ emission regions significantly overlap and that their respective ratios can be considered representative of the high excitation plasma. A concern, however, is that the weak \heiwar\ line lies very close to the \ariv\ \lam4711\AA\ line. Given the much wider profiles of the NLR lines in comparison to planetary nebulae, both lines will overlap. Hence the need to apply a deblending correction. The procedure we adopted for the single-density case is described in Appendix\,\ref{sec:ap-debhe}.  It essentially makes use of the dereddened \heiwr\ line (Col.\,6) to calculate a reliable estimate of the contribution of the \heiwar\ line to the blended\footnote{The sub-index $+$ sign denotes a line that incorporates a blended component from a different ion. The double  \lam\lam\ symbol refers to two separate but nearby lines of the same ion.}  {\arivp} \arivtop\  doublet ratio (Col.\,5). The estimated fractional contribution of the  blended \hei\ line to \arivp\ is \fblHe\ (Col.\,7), which amounts to $5$\% on average. NGC\,1068 stands out at a higher value of 10\%. For each object we iteratively determine  which density \nsng\ is implied by the deblended \ariv\ doublet ratio when it is calculated at the temperature \Tosng\ which reproduces the \Roiii\ ratio. The inferred density values, \nsng, given in Col.\,(8), all lie below $10^4$\,\cmc. As indicated by the cyan arrow in Fig\,\ref{fig:fig3}, the \Roiii\ ratio is not significantly affected by collisional deexcitation at densities below $10^4$\,\cmc. Taken at face values, the densities of Table\,\ref{tab:sam} indicate that \Roiii\ is a valid temperature indicator for the Seyfert\,2's of Kos78. The average temperature characterising the whole sample is $13\,980\pm 1200$\,\degk, which lies significantly above the predictions of the LDR photoionization models of \S\,\ref{sec:phot}. Rather than assuming a single density, in \S\,\ref{sec:isot} we will consider the case of a smoothly varying density distribution.

\subsection{Density bias due to a limited spatial resolution}
\label{sec:blur}

In Type\,I AGN, due to the favorable orientation of the observer with respect to the ionizing cone \citep{An93}, the densest NLR components are visible and possibly dominate the integrated line flux, causing the \Roiii\ ratios to occupy values up to 0.2 due to collisional deexcitation, as proposed by BL05.  

In Type\,II AGN on the other hand, since the inner regions  occupied by the accretion disk and the BLR are not visible, important selection effects take place. The NLR is likely not fully observed due to obscuration associated to the ionizing cone. 
Differences in spatial resolution as a result of the object distance and the size of the spectrograph aperture inevitably affect the sampling of the NLR volume. The angular resolutions characterising our Seyfert\,2 sample are the following. The reconstructed NLR spectrum of Ri14 was based on SDSS observations of Seyfert\,2's of similar redshifts (0.10 -- 0.12) with a fiber aperture size of 3\arcsec. This corresponds to a NLR sampling that extends over 5.6\,kpc diameter. The nearest AGN of the Kos78 sample is NGC\,1068 at $z=0.0038$, which is discussed in detail below. The other objects have redshifts in the range 0.0135 to 0.051 which, for an aperture of $2.7\arcsec \times 4\arcsec$, result in angular sizes in the range $\simeq 1$ to 4\,kpc at the object distance. The Be06b sample presents the highest spatial resolution since its four Seyfert\,2's are of low redshift (0.003 -- 0.027) and were observed with a seeing $\la 1$\arcsec using longslit spectroscopy mounted on the NTT and VLT telescopes of the European Southern Observatory. The slit aperture was $\simeq 1.1\arcsec \times 1\arcsec$, which translates into a NLR angular size of 50 to 600\,pc. We might conjecture that such superior spatial resolution is possibly related to the position of the Be06b sample in Fig.\,\ref{fig:fig1}, which is slightly more to the right than the Kos78 and Ri14 samples.

Unlike the  BL05 quasar sample where collisional deexcitation is the evident cause of the wide spread of \Roiii\ values, the NLR of Seyfert 2's appears relatively unaffected by deexcitation, at least among ground-based observations. The much superior resolution from HST observations, however, reveal the presence of  much denser components within the inner nucleus. A case in point are the HST-FOS observations of NGC\,1068 (Kr98) with an angular resolution of 0.3\arcsec\ (i.e.  25\,pc).
The nucleus (black open triangle), for instance, shows an \Roiii\ ratio higher by a factor two with respect to the ground-based observation of Kos78 (black octagon). The image of the nucleus in \oiii\ light using the HST-FOC instrument shows a diffuse underlying emission component that extends beyond 200\,pc from the nucleus and which encompasses a number of emission knots of sizes $\ga 10\,$pc.  Two bright (EELR)  emission `knots', labelled 1 and 2 by Kr98 (red filled triangles) observed with the 0.3\arcsec\ aperture show \Roiii\ values that fall between those of Kos78 (octagon) and of Kr98 with HST-FOS (open triangle). A red dotted line connects the four measurements in Fig.\,\ref{fig:fig1}. The two knots are situated at distances from the nucleus of 0.2 and 0.7\arcsec, respectively (i.e. at 16 and 57\,pc). 


\section{Multi-density temperature and density diagnostics}
\label{sec:isot}

We expanded the functionality of the \Roiii\ temperature \textit{diagnostic} by combining the latter with the \ariv\ density indicator, allowing us to evaluate the impact of collisional deexcitation among the observed  \Roiii\ ratios. To this effect, we developed the algorithm \OSALD\ (see App.\,\ref{sec:ap-osald} for further information), which integrates the line emissivities from an isothermal plasma that extends over a wide density range, up to a cut-off density, \ncut. At a given temperature, if \ncut\ has a high value, the integrated \Roiii\ ratio rises above the LDR value. Since our diagnostics depends on measurements of the \ariv\ doublet, we are limited to the Kos78 sample of Table\,\ref{tab:sam}. 

We have explored two options concerning the nature of the density cut-off: 1) that it simply consists of a sharp cut-off, or 2) corresponds to a gradual cut-off due to a foreground dust extinction layer whose opacity correlates with plasma density. The calculations assuming the first option are summarised in App.\,\ref{sec:ap-dfcase} and result in essentially the same temperatures as derived from the single density case explored in \S\,\ref{sec:sng}. We will now consider the second option where we explore the possibility that the NLR emission becomes gradually more absorbed towards the nucleus.

\subsection{Components of the dust screen approach with \OSALD} 
\label{sec:pldens}

One particularity of our proposed approach is that it implies fitting the \textit{observed} \Roiii\ ratios presented in Table\,\ref{tab:dust} rather than the dereddened ratios of  Table\,\ref{tab:sam} and Fig.\,\ref{fig:fig1}.

\subsubsection{A density cut-off generated by a dust extinction gradient} 
\label{sec:ingred}

The dust opacity is described by an exponential function of density $n$: $\tauv = \tauvi \exp(n/\nopa)$, where \nopa\ is the e-folding density that defines the gradual increase of the foreground \Vband\ dust opacity towards the inner nucleus. This definition does not require us to distinguish between the Galactic extinction from that from the NLR dust screen (\tauvi\ includes both). 
Our interest in exploring an ascending extinction towards the denser NLR component is motivated by the accumulating evidence of the importance of the orientation of the NLR (and not just of the BLR) with respect to the observer, as reviewed in App.\,\ref{sec:ap-orien}, and which is presumably the result of a cone-like opacity distribution.

\subsubsection{Extinction curve and line transfer algorithm }
\label{sec:exti}

The line transfer algorithm implemented in \OSALD\ fully takes into account the effect of multiple scattering across the foreground dust layers. Its characteristics are described in App.\,C of \citet{Bi93b}. As for the adopted extinction curve, we adopt the one inferred by \citet{MR91} in their study of the Orion nebula, which differs from the standard ISM curve in that grains of size smaller than 0.05\,$\mu$m are absent, resulting in a flatter curve with less extinction in the UV \citep{Ba91,Ma93}. It is qualitatively in line with the evidence presented by \citet{Ma02,Ma01} that small grains are depleted in the dusty medium which is responsible for the absorption of the X-rays and the reddening of the BLR lines.  The \Vband\ dust opacity \tauvi\ is determined by fitting the integrated  Balmer decrement, assuming recombination Case\,B at temperature \Toiii. The values of  \Toiii,  \nopa\ and \fblHe\ are set by iteratively fitting \Roiii\ and  the deblended \ariv\ \arivr\ ratio.

\subsubsection{Transposition to a simplified spherical geometry  }
\label{sec:trans}

The algorithm  consists in integrating the line emission measures\footnote{Defined as the line emission coefficient times the electronic density.} of an isothermal multi-density plasma (MDP) of temperature \temp. The calculations can be transposed to the idealized geometry of a spherical (or conical) distribution of ionization bounded clouds whose densities $n$ decrease as $r^{-2}$. The weight attributed to each plasma density component is set proportional to the \textit{covering solid angle}\footnote{$\Omega(n)= A(n)/4\pi r^2$ where $A$ is the area of a shell of density $n$ exposed to the ionizing source at a distance $r$. For definiteness we set the electron density equal to that of H, $n=n_e=n_H$.} $\Omega(n)$ subtended by the plasma shell of density $n$.  In the case of photoionization models, such a distribution would result in a constant ionization parameter \uout\ and the integrated columns $N_{\rm X_k}$ of each ion $k$ of any cloud would be to a first order constant. For the sake of simplicity, to describe $\Omega(n)$ we adopt the powerlaw $(n/\nlow)^{\epsi}$, which extends from $\nlow\ = 100\,$ up to $10^8$\,\cmc. If we transpose this to a spherical geometry where both \uout\ and $\Omega$ are constant (i.e. $\epsi=0$), the \textit{area} covered by ionization-bounded emission clouds would increase as $r^2$, thereby compensating the dilution of the ionizing flux and the density fall out (both $\propto r^{-2}$). In this case, the weight attributed by \OSALD\ to each shell is the same, otherwise  when $\epsi \ne 0$ the weight is simply proportional to $\Omega(n)$. MDP calculations are not a substitute to photoionization calculations. They are only intended as diagnostics that could constrain some of the many free parameters that characterize multidimensional NLR models, including those which might consider a non-uniform dust distribution.

\subsubsection{Selection of the distribution index $\epsi$ }
\label{sec:omeg}

To guide us in the selection of \epsi, we followed the work of Be06b who determined that, for a spectral slit radially positioned along the emission line cone, the surface brightness of the \textit{spatially resolved} ENLR is seen decreasing radially along the slit as $r^{\delta}$ (with $\delta < 0 $), where $r$ is the projected nuclear distance on the sky. From their \oiiiw\ and \ha\ line observations of Seyfert\,2's, Be06b derived average index values of $\delta_{[OIII]}=-2.24\pm 0.2$ and $\delta_{H\alpha}=-2.16\pm 0.2$, respectively. Let us assume that such gradient extends inward, i.e. inside the unresolved NLR. 
For our assumed spherical geometry where the \ha\ luminosity across concentric circular apertures behave as $r^{-2\epsi}$ (see \S\,\ref{sec:ap-o-subB}), $\epsi$ is given by $-(1+\delta)/2$. Hence we adopt $\epsi \approx +0.6$ in order that a long slit projected onto our spherical geometry could reproduce the observed $\delta_{[OIII]}$ value  of Be06b.  Out of curiosity, we have explored other positive values and found that changes in \epsi\ were not critical and did not affect our conclusions.

\subsubsection{Importance of deblending the \arivp\ \lam4711{\AA}{\tiny +} lines }
\label{sec:argon}

As emphasized by \citet{Ke19}, care must be taken when interpreting the \ariv\ doublet since the weak nearby \heiwar\ line is nearly superposed to the \ariv\ \lam4711\AA\ line, hence the need to apply a proper deblending correction to the measured \ariv\ ratios. The procedure adopted is described in App.\,\ref{sec:ap-debhe}. 
Because of the relative weakness of the \heiwar\ line, there is no direct evidence of its presence in AGN spectra given its closeness to the \ariv\ \ariva\ line. To deblend the flux contribution from the \heiwar\ line, we first evaluate its expected flux using the strong \heiwr\ line and then subtract it from the \ariv\ \lam4711\AA\ line.  Since we are dealing with He recombinations lines, the dependence of the \hei\ ratio on temperature or density is relatively minor. Hence obtaining a reliable estimate of the \hei\ fractional contribution, \fblHe, to the observed \ariv\ profile is straightforward. The procedure is described in Appendices\,\ref{sec:ap-debhe} and \ref{sec:ap-o-subD}.

Another potential blending consists of the first two lines of the \neiv\ quadruplet which comprise the lines \lam4714.36, \lam4715.80, \lam4724.15 and \lam4726.62\AA\  \citep{GR15}. For convenience, we will refer to them as consisting of two doublets centered at \lam\lam4715\AA\ and \lam\lam4725\AA, respectively. Up to densities of $\sim 10^6\,$\cmc, the unblended \neivb\ doublet is calculated to be on average 35\% brighter than the \neiva\ doublet. In those cases where the \neivb\ doublet is detected, we can reliably determine the blended \neiva\ doublet flux and then subtract it from the blended \arivp\ \lam4711\AA{\tiny +} lines. Further information about the deblending procedure is given in App.\,\ref{sec:ap-debne}. No detection of the \neiv\ \neivb\ line was reported by Kos78.


\begin{figure}[!t]
  \includegraphics[width=\columnwidth]{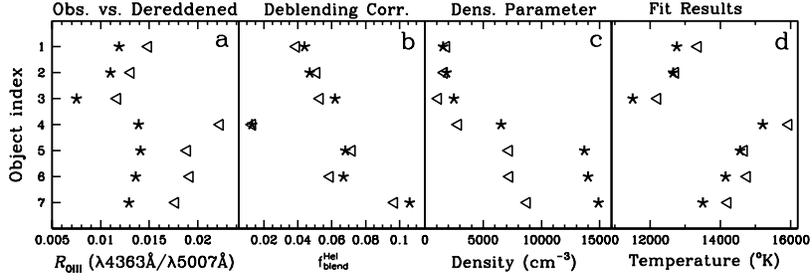}
  \caption{Comparison between parameters and plasma temperatures shown in Table\,\ref{tab:sam} (open triangles) and Table\,\ref{tab:dust} (stars). An index number (1 -- 7) identifies the object name in Col.\,(2) of either Table. Panel a: target \Roiii\ ratios, dereddened (open triangles) vs. observed (stars). Panel~b: deblending corrections \fblHe\ applied to the \neiv\ doublet ratios. Panel\,c: sharp density cut-off (open triangles) vs.  extinction density cut-off \nopa\ (stars). Panel\,d: plasma temperatures inferred from the dustfree model (open triangles) vs. dusty screen model (stars).
   }
\label{fig:fig4}
\end{figure}

\subsection{Results from the multi-density plasma models  }
\label{sec:fit}

\subsubsection{The reference Seyfert\,2 sample}
\label{sec:noneiv}


\begin{table}
\caption{OSALD parameter fit of observed line ratios\tabnotemark{a}}
\label{tab:dust}
\begin{changemargin}{-0.5cm}{-0.5cm}
\centering
\small
    \begin{tabular}{cl ccc cc cccc}
      \toprule
 \multicolumn{2}{c}{Objects}  & \multicolumn{3}{c}{Target line ratios}  & \multicolumn{3}{c}{Blending corrections}  & \multicolumn{3}{c}{Parameter values }\\
 \cmidrule(r){1-2} \cmidrule(lr){3-5} \cmidrule(lr){6-8} \cmidrule(lr){9-11} 

\hspace{0.015cm} (1) & \hspace{0.025cm} (2) & (3) & (4) & (5) & (6) & (7) & (8) & (9) & (10) & (11)\tabnotemark{b} \\
\multicolumn{1}{c}{Index} & {AGN} & \ha/\hb & \Roiii & \ariv{\tiny +}  &  \RHeAr\  & \fblHe & \ariv & \large{\tauvi} & \large{\nopa}  &  \Toiii  \\

\# &  & $\frac{\lamf4861}{\lamf6563}$ & $\frac{\lamf4363}{\lamf5007}$ & $\frac{\lamf4711{\tiny +}}{\lamf4740}$ & $\frac{\lamf5876}{\lamf4740}$ &  & $\frac{\lamf4711}{\lamf4740}$ &  & \cmc  & \degk  \\ 
 \midrule
{1} & {Mrk\,573}  & {3.62} & {0.0119} & {1.156}  & {2.03} & {0.044} & 1.108 & {0.16} & {$1.56\times 10^3$} & {12\,760} \\

{2} & {Mrk\,34}  & {4.10} & {0.0110} & {1.193}  & {2.46} & {0.047} & 1.140 & {0.41} & {$1.85\times 10^3$} & {12\,660} \\

{3} & {Mrk\,78}  & {5.31} & {0.0075} & {1.238}  & {4.05} & {0.062} & 1.166 & {1.10} & {$2.48\times 10^3$} & {11\,510} \\

{4} & {Mrk\,176}  & {6.55} & {0.0139} & {1.045}  & {0.90} & {0.013} & 1.031 & {1.74} & {$6.54\times 10^3$} & {15\,210}  \\

{5} & {Mrk\,3}  & {5.31} & {0.0141} & {0.850}  & {3.00} & {0.068} & 0.796 & {1.08} & {$1.37\times 10^4$} & {14\,560}  \\

{6} & {Mrk\,1}  & {5.00} & {0.0136} & {0.814} & {2.71} & {0.067} & 0.762 & {0.91} & {$1.40\times 10^4$} & {14\,150} \\

{7} & {NGC\,1068}  & {4.47} & {0.0129} & {0.763}  & {3.56} & {0.106} & 0.690 & {0.61} & {$1.49\times 10^4$} & {13\,500} \\

      \bottomrule
    \end{tabular}
    \begin{tablenotes}
      \small
      \item $^a$ The line ratios were not corrected for reddening. A foreground dust screen was assumed instead whose opacity increases exponentially: $\tau_{ij} = \tauvi \exp(n/\nopa) \, A(\lambda_{ij})/A_{V}$, where $A(\lambda_{ij})/A_{V}$ represents the extinction curve evaluated at wavelength $\lambda_{ij}$ for the emission line  $ij$ considered.  
      \item $^b$ The averaged temperature for the sample is $\Tmean = 13\,480\pm 1180$\,\degk. 
    \end{tablenotes}
  \end{changemargin}
\end{table}

The results from the calculations using \OSALD\ are presented in Table\,\ref{tab:dust} where we have assumed a powerlaw index of $\epsi=+0.6$. In Col.\,(7), \fblHe\  represents the estimated blending contribution from \hei\ \lam4713\AA\ to the \arivp\ \lam4711\AA{\tiny +} lines. The resulting deblended \ariv\ \arivr\ ratios are presented in Col.\,(8). The foreground dust screen opacities, \tauvi, inferred  by \OSALD\  from the observed Balmer decrements (Col.\,3) are given in Col.\,(9). The fitted dust distribution e-folding densities, \nopa, and the inferred plasma temperatures are given in Cols.\,(10) and (11), respectively.

To facilitate the evaluation of these fits, we show in Fig.\,\ref{fig:fig4} how the parameters compare for each object between Table\,\ref{tab:sam} (triangles) and Table\,\ref{tab:dust} (stars). Panel\,(a) represents the target \Roiii\ ratios, dereddened vs. observed, panel\,(b) the deblending correction \fblHe, panel\,(c) the the cut-off density vs. dust drop-out density scale, and panel\,(d) the temperatures inferred from the fits. The e-folding dust screen densities, \nopa, derived by OSALD lie in the range 1500 to 16\,000\,\cmc. The average \Tmean\ for the seven Seyfert\,2's is $13\,480\pm 1180$\,\degk, being lower by only 500\,\degk\ with respect to the single density case. 

Having considered an explicit density distribution, with either dust obscuration (Col.\,10) or without (see Table\,\ref{tab:pld} in App.\,\ref{sec:ap-dfcase}), we derive temperature values that do not differ much from the single density case of Table\,\ref{tab:sam}. This supports our contention that collisional deexcitation is not affecting significantly the \Roiii\ ratios observed by Kos78. Given the relative proximity in Fig.\,\ref{fig:fig3}  of the  Ri14 subset a41 (black diamond) to the Kos78 sample (black disk), we might conjecture that LDR possibly applies to the a41 sample as well since, at a redshift of $z=0.11$, the large projected scale of the 3\arcsec\ SDSS aperture ensures significantly more dilution of the inner dense NLR component in comparison to the Be06b and Kos78 samples. We would need a larger sample of \ariv\ doublet measurements to confirm that \Roiii\ translates into a reliable determination of the plasma temperature in Type\,II AGN.

\subsubsection{Probing the possible blending of \ariv\ by \neiv}
\label{sec:kosneiv}

We present complementary calculations in Table\,\ref{tab:dustneon} where we have assumed the hypothetical case of the \neivb\ doublet reaching 30\% of the observed \ariv\ \arivb\ line intensity. The blending contribution from the \hei\ and \neiv\ \neiva\ lines, that is \fblHe\ and \fblNe, are presented in Cols.\,(7) and (9), respectively, and the resulting deblended \ariv\ doublet ratios are listed in Col.\,(10). The opacities \tauvi\ (Col.\,11) inferred remain about the same, but the e-folding densities \nopa\ (Col.\,12) are typically larger by a factor with respect to Table\,\ref{tab:dust}. The average sample temperature \Tmean\ is lower by only 125\,\degk. At least for the Kos78 sample at hand, not including the \neivb\ doublet should not affect our conclusions concerning the Seyfert\,2 NLR temperatures. 

\subsubsection{Detection of \neiv\ in nearby Seyfert\,1 NGC\,4151}
\label{sec:evaneiv}

The critical densities of the \neiv\ quadruplet lines all lie above $10^6$\,\cmc. Because the \ariv\ doublet emissivities at such densities are significantly reduced due to collisional deexcitation, a positive detection of the \neiv\ \lam\lam4725\AA\ doublet might relate with having plasma densities much above those deduced from the Kos78 sample. This might be the case in Type\,I AGN. Interestingly, the detection of the \neiv\ \neivb\ doublet was reported early on in the Seyfert\,1 NGC\,4151 by \citep{Bok75}. The line ratios of interest for this object are shown in Table\,\ref{tab:dustneon}. A labeled star depicts its position in Fig.\,\ref{fig:fig3}. Eye estimates of the \neiv\ \lam\lam4725\AA\ doublet (from the published figure) suggest a value of $\sim 0.3$ with respect to the \ariv\ \arivb\ line while the measurements reported in their Table\,I  would imply a higher value of $0.62$. In Table\,\ref{tab:dustneon}, we present three \OSALD\ fits in which the \neiv/\ariv\ ratio (Col.\,8) successively takes on the values of 0, 0.3 and 0.62. The two \neiv\ deblending corrections result in  \nopa\ values higher by factors of  2.2 and 7.5 for models 8B and 8C, respectively, with  the deblended \ariv\ doublet ratios dropping to 0.496 and 0.294 (Col.\,10). The impact on the inferred temperature is significant, with \Toiii\ from model 8C being 1870\,\degk\ lower, at 16\,180\,\degk, showing minimal evidence of collisional deexcitation being present. It would be interesting to repeat this exercise if we could obtain higher S/N spectra.

\subsubsection{The particular case of QSO\,2's}
\label{sec:qsotwo}

Through our literature search of Type\,II AGN measurements of the \lam\lam4725\AA\ \neiv\ doublet, we came across three objects classified as QSO\,2's, that is, Type\,II quasars corresponding to the high luminosity counterpart of Seyfert\,2's. They are Mrk\,477 \citep{VM15}, SDSS J1300+54 and SDSS J1653+23 \citep{VM17} at redshifts $z$ of 0.037, 0.088 and 0.103, respectively. Their spectra were extracted from the Sloan Digital Sky Survey data \citep[SDSS;][]{Yo00} and the line ratios relevant to our analysis are given in Table\,\ref{tab:dustneon}. 

What stands out from these objects is their larger \Roiii\ ratios. The deblended \ariv\ doublet ratios (Col.\,8) do not imply significant collisional deexcitation, except at a reduced level in Mrk\,477 where \nopa\ reaches $\simeq 22\,200\,$\cmc. Yet the  \Toiii\ values inferred (Col.\,13) for the three QSO\,2's are much higher than in Seyfert\,2's, which questions the plausibility of LDR conditions. It is possible that AGN where the \neiv\ \neivb\ can be detected might indicate the presence of a double-bump in their density distribution.  We tentatively explored the addition of an additional denser plasma component ($\ga 10^6$\,\cmc) to our powerlaw. Our fit to the \ariv\ doublet was not very sensitive to this component since  both \arivl\ lines are affected by collisional deexcitation at the high density end. Even though the temperatures we inferred came out at values lower than in Col.\,(13), the exercise was not convincing as the number of free parameters exceeded the number of variables. A possible interpretation is that the super-luminous QSO\,2's scale up in size to the extent that their inner NLR become partly visible as is the case with the high spatial resolution HST measurement of the nearby Seyfert\,2 NGC\,1068 (c.f. red triangles in Fig.\,\ref{fig:fig1}). 

\section{Temperature problem with photoionization}
\label{sec:tepr}

In conclusion, after integrating the emissivities of the \oiii\ and \ariv\ lines over a continuous distribution of densities, we find that the impact of collisional deexcitation on the \lam4363\AA/\lam5007\AA\ (\Roiii) ratio is not significant among ground-based observations of the seven Seyfert\,2 sample of Kos78 who provided measurements of the \ariv\ density  indicator, and therefore, their \Roiii\ ratio provides us with a reliable measurement of the NLR temperature. A comparison of the values of \Roiii\ observed among quasars, Seyfert\,2's and spatially extended ENLR plasma, as displayed in Fig.\,\ref{fig:fig1}, argues in favor of a floor temperature for the NLR, which we situate at $\ga 13\,500$\,\degk. Our photoionization models using \map\ and assuming standard SEDs and low densities predict \Roiii\ values significantly below those observed in Seyfert\,2's. This discrepancy defines what we would label the {\Roiii}-temperature problem. 

In the current work, we found complementary evidence that the orientation of the emission line cone  with respect to the observer's line-of-sight affects our characterisation of the NLR, whether in Seyfert\,2's (Kos78 sample) or in quasars (BL05 and RA00 samples). We do not exclude the existence of a much denser NLR component being present in ground-based observations, but we would propose that, at least among Seyfert\,2's with $z \ga 0.02$, the latter would be strongly diluted by the much brighter low density NLR component, which we evaluate has a density $\la 10^{4.3}\,$\cmc.  In quasars, where a larger fraction of this dense and luminous component becomes visible, the resulting \Roiii\ ratio progressively increases up to values of $\sim 0.2$ due to collisional deexcitation, as proposed by BL05 using dual-density photoionization models. It would be interesting to investigate whether the \neiv\ \lam4725\AA\ doublet becomes intrinsically stronger in Type\,I objects. A few luminous AGN in which we reported the detection of the \neiv\ doublet in \S\,\ref{sec:evaneiv} and \ref{sec:qsotwo} appear to favor a density distribution akin to a double-bump, such as the dual-density approach of BL05, rather than the single powerlaw we have assumed. 

\begin{acknowledgements}
This work has been partly funded with support from the Spanish Ministerio de Econom\'{\i}a y Competitividad through the grant AYA2012-32295. G.MC is grateful for the support from the Centro de Investigaciones de Astronom\'{\i}a (CIDA). A.R.A acknowledges partial support from CNPq Fellowship (312036/2019-1 and 203746/2017-1). M.M-P acknowledges support by the Korea Astronomy and Space Science Institute for her postdoctoral fellowships. A. Alarie was funded by a postdoctoral grant from CONACyT. We thank the referee for his helpful constructive comments. 

The discussion about the referred three QSO\,2's is based on data from the Sloan Digital Sky Survey. Funding for the SDSS and SDSS-II has been provided by the Alfred P. Sloan Foundation, the Participating Institutions, the National Science Foundation, the US Department of Energy, the National Aeronautics and Space Administration, the Japanese Monbukagakusho, the Max Planck Society and the Higher Education Funding Council for England. The SDSS website is http://www.sdss.org/. The SDSS is managed by the Astrophysical Research Consortium for the Participating Institutions.
\end{acknowledgements}

\begin{appendices} 



\section{The validity of low density regime for the ENLR}
\label{sec:ap-enlr}

Among the Seyfert galaxies studied \citep[e.g.][]{Be06a,Be06b}, the ENLR densities inferred from the red \siiw\ doublet are typically $< 10^3\,$\cmc. Along the long-slit measurements, in most cases both the electron density and the ionisation parameter appear to be decreasing with radius. The authors proposed that deviations from this general behaviour (such as a secondary peak), when seen in both the ionisation parameter and electron density space, can be interpreted as signs of shocks due to the interaction with a radio jet.
In what follows, we will consider those cases where the excitation mechanism is photoionization by the accretion disk. The red dot in Fig.\,\ref{fig:fig1} represents the average \Roiii\ of the ENLR of four\footnote{Each of these measurements is associated to a Seyfert\,2 as we left out the fifth object, NGC\,526A, which is classified Seyfert\,1.5.} Seyfert\,2's observed by SB96. 
If we assume a temperature of 9\,000\,\degk\ for the \sii\ plasma, the densities inferred by BWS from the \siirr\ ratios from each ENLR are $\le 250$\,\cms. Two other examples shown in Fig\,\ref{fig:fig1} are: a) the yellow filled dot, which corresponds to a deep spectrum of a 8\,kpc distant cloud from the nucleus of radio-galaxy Pks\,2152-699 (Ta87), and b) the magenta dot which represents the average \Roiii\ of seven optical filaments situated at $\sim 490$\,pc from the nucleus of the radio-galaxy Centaurus\,A (Mo91).  In both cases, the red \sii\ doublet measurements indicate low electronic densities, with $n_e \le 250$\,\cmc. 

High excitation ENLR lines such as \oiii\ should similarly operate under LDR conditions for the following reasons. First, the geometrical dilution factor of the ionizing flux across the typical detector aperture does not vary significantly. For instance, if we define $r_{in}$ as the radial distance separating the UV source from the inner boundary of the observed ENLR, and $\Delta r$ the radial thickness corresponding to the detector aperture projected on the sky, then the ratio $(\Delta r/r_{in})^2$  represents the fractional change of the UV dilution factor across the observed ENLR. This factor is typically $\la 1.5$, indicating  that the observed plasma is exposed essentially to the same ionizing flux. Line ratio variations across $\Delta r$ must be due to either variations in plasma density or to a progressive absorption of the ionizing radiation, but not to changes in the dilution factor. Second, if the \oiii\ and \sii\ lines originates from unrelated gas components, the densities of the low \sii\ emission plasma must be order of magnitudes \textit{higher} than the \oiii\ emission plasma in order to sufficiently reduce the ionization parameter to the point that the low ionization species dominate the spectrum. Third, if the \sii\ emission corresponds to the partially ionized layer at the back of a photoionized (ionization-bounded) slab, the electronic density of the region that emits the \oiii\ lines is denser by a factor of 2.5 to 4. The reason is that the \sii\ emission comes from plasma that is partially ionized with an electronic density $n_e$ much lower than the local gas density $n_H$.  This factor is sufficiently small that for the typical \sii\ densities of $n_e \simeq 250\,$\cmc\ as found in the ENLR, the density associated to the  \oiii\ lines would still be $\ll 10^4\,$\cmc\ and LDR conditions should therefore apply.


\section{Correcting the \ariv\ doublet from line blending}
\label{sec:ap-deblend}

\subsection{Correcting the \arivp\ ratio from \hei\ \lam4713\AA\ blending}
\label{sec:ap-debhe}

One characteristic of ratios involving recombination lines of the same ion is their limited sensitivity to either temperature or density. A satisfactory prediction of the \hei\ \lam4713\AA\ line intensity can be derived from the measurement of the \hei\ \lam5876\AA\ line.   
First, we derive the Case\,B \hei\ \lam4713\AA/\lam5876\AA\ ratio, which we label \Rhei, via interpolation of the emissivities from the supplemental Table of \citet{Po13}. For a  $10^4\,$\cmc\ plasma at a temperature of 12\,000\,K, \Rhei\ turns out to be only 4.78\%  of \hei\ 5876\AA. Temperature variations of $\pm 2000$\,\degk\ would cause a change in this ratio of $^{+7.95}_{-11.5}$\%, respectively, while adopting density values of 100 and 10$^5\,$\cmc\ would result in \Rhei\ ratios of 0.0429 and 0.0489, respectively.
Second, by defining \RHeAr\ as the observed \hei/\ariv\ \lam5876\AA/\lam4740\AA\ ratio, the product $(\RHeAr \times \Rhei)$ defines our estimate of the blending contribution from \hei\ \lam4713\AA\  to the \textit{measured} (blended) {\arivp} doublet ratio. The relevant information is provided by the fractional contribution of the blended line, which is given by $\fblHe= (\RHeAr \times \Rhei)/\arivp$. The blended-corrected \ariv\ \arivr\ doublet ratio is given by $(1-\fblHe) \times \arivp$, which was used to explore the single density case discussed in \S\,\ref{sec:sng} (see Col.\,(7) of Table\,\ref{tab:sam}). For the powerlaw density distribution case, the procedure is described in \S\,\ref{sec:ap-o-subD}.

\subsection{Correcting the \arivp\ ratio from \neiv\ \lam\lam4715\AA\ blending}
\label{sec:ap-debne}

The \neiv\ optical lines consist of a quadruplet at \lam4714.36, \lam4715.80, \lam4724.15 and \lam4726.62\AA, respectively  \citep{GR15}. To simplify the notation, we will refer to the quadruplet as consisting of two doublets: the observed \neiv\  \neivb\ lines and the potentially blended \neiv\ \neiva\ lines. At typical NLR densities, the potentially observed \neiv\ \neivb\ doublet is calculated to be $\approx 35$\% brighter than the \neiv\ \neiva\ doublet (blended with \ariv\ \lam4711\AA). To our knowledge the \neiv\ \neivb\ doublet has only been reported in a few AGN. However, it is frequently observed in planetary nebulae (PN). For instance, in NGC\,6302 where the stellar temperature is estimated to be in the range 224\,000 to 450\,000\,\degk\ \citep{Fe01}, the observed \neiv/\ariv\ \neivb/\arivb\ ratio reported by \citet{AL81} is 0.175. Since the emission lines of PN are narrow, the above mentioned lines can be resolved without any need of deblending provided high resolution spectroscopy has been carried out. For instance, the observations of the PN NGC\,3918 by \citet{GR15} using the Ultraviolet-Visual Echelle Spectrograph (UVES, \citet{DO00}) with an 1\arcsec\ slit resulted in a spectral resolution of 6.5\,\kms. The observed emission profiles of each \neiv\  quadruplet line were well resolved, showing a FWHM of $\simeq 20\,$\kms. Interestingly, the \neiv/\ariv\ \neiva/\ariva\ and \hei/\ariv\ \lam4713\AA/\lam4711\AA\ ratios measured by \citet{GR15} are 0.115 and 0.078, respectively. To correct the measured \arivp\  (\arivtop)  ratio from \neiv\ blending, the density integration by \OSALD\ of the two doublets \neiv\  \neivbb\ and  \neiv\ \neivaa\ proceeds as described in App.\,\ref{sec:ap-o-subD} for the \hei\ \lam4713\AA\ and \lam5876\AA\ lines.


\section{The OSALD algorithm} 
\label{sec:ap-osald} 

Using emission line atomic physics, \OSALD\footnote{Stands for "Oxygen Sulfur Argon Line Diagnostic".} explores temperature and density diagnostics in which an explicit distribution of the density is considered.  

\subsection{Line diagnostics with a powerlaw density distribution}
\label{sec:ap-o-subA}

Our goal is to explore which density distribution best reproduce a given set of line ratios and to determine to what extent collisional deexcitation is affecting the observed \Roiii\ \oiiir\ or \Rnii\ \niir\ line ratios. For the high ionization species such as \opp, \arppp\ and \neppp, the temperature \Toiii\ is set iteratively to the value which reproduces the target \Roiii\ ratio. Although not considered in the current Paper, other diagnostics can be modeled such as the singly ionized oxygen \oii\ \oiirr\ and \oii\ \oiibr\ line ratios at the temperature that would reproduce the temperature sensitive  \Rnii\ target ratio, or the singly ionized sulphur \sii\ \siirr\ and \sii\ \siibr\ line ratios at the estimated temperature $\Tsii \sim 9000\,$\degk.

\subsection{Transposition of OSALD to a spherical geometry}
\label{sec:ap-o-subB} 

The isothermal plasma considered by \OSALD\ can be visualized as consisting of  concentric shells of plasma whose densities decrease radially as $r^{-2}$. These shells are given a weight which we associate to the covering solid angle of a putative ionizing source at the center. This can be transposed to the idealized case of photoionized shells that are ionization bounded and share the same ionization parameter \uout. To the extent that the low density regime applies, the line luminosities of each shell result equal if they share the same covering \textit{solid angle} $\Omega$ of the ionizing source. \OSALD\ basically integrates the line emission coefficient times the shell covering solid angle: $j_{ij}^k(n,T)\, \Omega(n) \, {\Delta}n$, where $ij$ corresponds to the transition from level $j$ to $i$ evaluated at temperature $T$ and density $n$, which takes into account collisional deexcitation. 
The integrated line flux for line $ij$ of ion $k$ reduces to the summation in density space of  $\sum_{l} \,\Omega(n_l) ~ j_{ij}^k(T,n_l) \ {\Delta}n_l$, where $n_l$ is progressively increased in locked steps of size ${\Delta}n_l/n_l = 0.004$\,dex from $\nout = 100$ up to the cut-off density \ncut. When the fit incorporates foreground dust extinction, \ncut\ is fixed at  ${10}^8\,$\cmc\ and the actual cut-off is set by the foreground dust extinction which increases exponentially with density. The e-folding density for the opacity in the V-band is defined by the parameter \nopa (see \ref{sec:ap-o-subC}). The weight of each shell is given by its covering solid angle $\Omega$, which follows a powerlaw of index \epsi\ with density: $\Omega(n) = \Omega_o ~ (n/\nout)^{\epsi}$, where ${\Omega_o}$ is an arbitrarily small constant that would ensure negligible shell shadowing. The density is postulated to decrease radially as $n \propto r^{-2}$. 
As a result, the shells' luminosities behave as $r^{-2\epsi}$ since positive values of \epsi\ imply a covering solid angle that increases towards the ionizing source (along with the density $n$). A slit radially positioned along the ionizing cone would result in  an \hb\ surface brightness that decreases as $r^{\delta}$, with $\delta = -(2\epsi+1)$.

\subsection{Line transfer across the cone-like dust screen} 
\label{sec:ap-o-subC} 

In the context of Type\,II objects, we propose in \S\,\ref{sec:isot} that each emission line is seen through a dust screen whose opacity increases exponentially towards the inner denser regions. For each line $ij$, the opacity $\tau_{ij}(n)$ is given by $\tauvi \exp(n/\nopa) \, A(\lambda_{ij})/A_{V}$, where \tauvi\ is the \Vband\ dust opacity at the lowest density \nout, \nopa\ is the e-folding density of the exponential function, $A(\lambda_{ij})$ the selected extinction curve and $A_{V}$ the extinction value at 5\,500\AA. When integrating the emission measures, the dust transfer function ${\rm Tr}(\tau_{ij}(n))$ is applied to each emission coefficient $j_{ij}^k(n,T)$. The latter assumes a plane-parallel geometry and takes into account both absorption and scattering due to the dust grains, as described in App.\,C of \citet{Bi93b}. In order to constrain the parameter \tauvi, the set of line ratios that are selected to be fitted must include one or more Balmer line ratios from H.

\subsection{Blending of the \hei\ \lam4713\AA\ and \ariv\ \lam4711\AA\ lines} 
\label{sec:ap-o-subD} 

To calculate the flux of any \hei\ line, the Case\,B recombination coefficients are taken from the work of \citet{Po13}. They cover the temperature range $5\,000 \le T_{rec} \le 25\,000\,$\degk\ and density range $10^2 \le n_e \le 10^{14}\,$\cmc.  By default the temperature assumed for \hei\ in the current work is  $T_{rec}= 12\,000$\,\degk\ while it is the variable \Toiii\ for \ariv\ and all the high ionization ions. 

In order to evaluate the blending of the weak \hei\ \lam4713\AA\ line with the \ariv\ \lam4711\AA\ line,  \OSALD\ integrates the emission flux of the following four lines:  \hei\ \lam4713\AA, \hei\ \lam5876\AA, \ariv\ \lam4711\AA\ and \ariv\ \lam4740\AA, taking into account dust extinction at the corresponding densities. This procedure properly takes into account how the emission coefficient of each line is affected by density and collisional deexcitation as well as by dust extinction which may increase along with density. After assuming an arbitrary abundance ratio of the two ionic species \hep\ and \arppp, the  algorithm derives the integrated \hei/\ariv\ ratio labelled $R_{5876/4740}$ and rescale it to the observed value. The $R_{4713/4740}$ ratio represents a measure of the blending contribution from \hei\ and is derived from the ratio $R_{4713/5876}/R_{5876/4740}$.   Deblending the \ariv\ $R_{4711/4740}$ ratio is achieved by subtracting the $R_{4713/4740}$ ratio from the observed blended \arivp\ ratio.  The fraction of \arivp\ due to \hei\ blending is labelled \fblHe\ in all our Tables.

\subsection{Minimum \chin\ and iterative least squares fit} 
\label{sec:ap-o-subE} 

We used a non-linear least squares fit method to find the optimal input parameter values that succeed in reproducing as closely as possible the target line ratios. These parameters are  varied in iterative fashion until the minimum re-normalized \chin\ value is encountered, with \chin\ defined as
\begin{equation}
 \chin(x_j) = \sum_{i=1}^{m} 
 \frac{w_i \, \big(y_i-y(x_j)\big)^2}{MAX\big[y_i^2 \, , \, y(x_j)^2\big]} \biggm/ \sum_{i=1}^{m} w_i
\end{equation}

\noindent where $m$ is the number of line ratios simultaneously fitted, $w_i$ the weight attributed to each line ratio $i$, $y_i$ the observed target line ratios and $y(x_j)$ the corresponding line ratios derived from the integration of the line fluxes. The quantity $x_j$ represents the various parameters on which the line integration depends on, that is, the temperature $T_{fit}$ and density $n$ as well as the parameters describing the behaviour of the covering angle $\Omega(n)$, which are \epsi\ and \ncut\ as defined in App.\,\ref{sec:ap-o-subB}.  As detailed in \S\,\ref{sec:fit} and Table\,\ref{tab:dust}, we used the algorithm to fit the line ratios of the seven Type\,II NLR of Table\,\ref{tab:sam}. By trial and error we settled for weights $w_i$ of 2.0 and 1.5 for the \ariv\ \arivr\ and \Roiii\ \oiiir\ ratios, respectively.

\OSALD's basic goal is to evaluate whether or not there is evidence of significant collisional deexcitation affecting the \Roiii\ ratio of any AGN whose \arivr\ ratio is successfully measured. Any fit where \chin\ exceeds 0.05 is deemed unsatisfactory and of no use. The fits described in Tables\,1--3 all present a negligible \chin\ $\sim 5 \times 10^{-7}$. For this reason, the line ratios derived  from the fits are, for all practical purposes, equal to the target ratios.


\section{NLR orientation and the observer's perspective} 
\label{sec:ap-orien} 

The geometrical set up behind the \textit{unified model} may apply not only to the BLR but to parts of the NLR that are gradually obscured in Type\,II objects. This would explain why the NLR line emission observed in the Seyfert\,2's correspond to a much lower density plasma than observed in Type\,I's. Examples of studies confirming the impact of the observer's perspective on the NLR are:

\begin{list}{}{}

\item[a)]  Using a data set of 18 Seyfert\,1 and 17 Seyfert\,2 of similar redshift from the literature, \citet{Mu98a} showed the evidence of an excess of \feviiw\ emission in Type\,I AGN with respect to Type\,II. The \fevii/\oiii\ (\lam6087\AA/\lam5007\AA) ratio in Type\,I AGN turns out to be an order of magnitude larger than in Type\,II. They proposed that it was linked to a region residing in the inner wall of a dusty torus, which they labeled the high-ionization nuclear emission line region (HINER\footnote{Terminology suggested by \citet{Mu98c} to contrast those AGN from LINER.}). 

\item[b)]Using a sample of 214 Seyferts, \citet[hereafter NMT]{NMT} confirmed that Type\,I Seyferts show a statistically higher \Roiii\ than Type II Seyferts. 
Using the work of \citet{DERO84,DERO86} who measured the line widths of 24 Seyferts, MNT found that the FWHM of \oiiitw\ in Type\,I spectra was larger than that of \oiiiw\ while in Type\,II spectra the FWHM of both lines were statistically indistinguishable. Although with less statistical significance, two more results were presented by NMT: 1) the FWHM of \oiiitw\ was larger in Type\,I than in Type\,II spectra, and 2) the FWHM of \oiiiw\ in Type\,I and Type\,II spectra were statistically indistinguishable. The authors commented that these results would suggest that the  strongly \oiiitw\ emitting region is located in a deeper inner region as compared to \oiiiw\ and that it is fully visible only in Type\,I AGN.  MNT inferred that the dependence of \Roiii\ on AGN types can be attributed to obscuration effects. 

\item[c)] \citet[]{Me08} favor a similar interpretation with respect to the mid-infrared coronal lines. They found that the mean \oiiiw\ line luminosity is 1.4\,dex smaller in Seyfert\,2's than in Seyfert\,1's while in the case of the mean \oivw\ line luminosity, the difference between the two subgroups is only 0.2\,dex. Their linear regression in the log plane of each AGN subgroup reveals that the luminosity of \oiii\ scaled almost linearly as $\Loiv^{0.9\pm 0.1}$ in Seyfert\,1's, but much more steeply, as $\Loiv^{1.8\pm 0.5}$ in Seyfert\,2's. Both trends are consistent with strong dust absorption of \oiii\ while \oiv\ is relatively little affected by extinction. 
It confirms earlier reports of \citet{JB91,CAM93,Mu94,Ke94,RL05,Nz06} that a much higher dust extinction affects the optical NLR of Seyfert\,2's than of Seyfert\,1. 

\item[d)] Finally, the work of \citet{Ros15a,Ros15b} of Coronal-Line Forest Active Galactic Nuclei  (CLiF AGN), which are characterized by a rich spectrum of optical forbidden high-ionization lines, suggests that the inner obscuring torus wall is the most likely location of the coronal line region. 

\end{list}


\section{Dust-free OSALD calculations} 
\label{sec:ap-dfcase} 

In \S\,\ref{sec:sng}, we assumed a single plasma density to determine the plasma temperature of the Kos78 Seyfert\,2 sample. We present in Table\,\ref{tab:pld} calculations from \OSALD\  where the density is represented by a powerlaw density distribution of index $\epsi=+0.6$ that extends from $n=100$\,\cmc\ up to a sharp cut-off density \ncut. The blending corrected  \ariv\ \arivr\ ratios are given in Col.\,(4) where we followed the procedure described in Appendices\,\ref{sec:ap-debhe} and \ref{sec:ap-o-subD} to evaluate the \hei\ \lam4713\AA\ fractional contribution \fblHe\ to the blended \arivp\ line.  The free parameters \ncut\ and \Toiii\ were iteratively varied until they reproduced the dereddened ratios of both \Roiii\ (Col.\,4) and (blended) \arivp\  (Col.\,5) from data Table\,\ref{tab:sam}. The  inferred values for \ncut\ and \Toiii\  are given in Cols.\,(5--6) of Table\,\ref{tab:pld}. 

The plasma temperatures \Toiii\ of Table\,\ref{tab:pld} are essentially the same as those of Col.\,(9) from Table\,\ref{tab:sam} that were derived assuming a single density. The main reason is that, insofar as the \oiii\ lines are concerned, the line emissivities for the whole sample take place in the low density regime and, as a result, the density averaged over the whole distribution, $\bar n$, turns out to be very close to the single density value \nsng\ from Table\,\ref{tab:sam} (Col.\,8). For instance, for the object with the highest cut-off density of Col.\,(5), NGC\,1068, the ratio \Roiii\ increases by only 3\% across the range of densities covered by the powerlaw distribution and the average density, $\bar n$, is 9425\,\cmc, which is close to the single density \nsng\ value of 8770\,\cmc. For the four objects where $\ncut < 10^4\,$\cmc, the mean densities $\bar n$ are proportionally closer to the corresponding \nsng\ values.

\setcounter{table}{0}
\renewcommand{\thetable}{B\arabic{table}}
\begin{table}[!t]
\caption{OSALD dustfree fit of dereddened ratios\tabnotemark{a} }
\label{tab:pld}
\begin{changemargin}{-0.5cm}{-0.5cm}
\centering
\small
\begin{tabular}{clcccc}
\toprule
\hspace{0.015cm} (1) & \hspace{0.225cm} (2) & (3) & \hspace{0.125cm} (4)\tabnotemark{b} & \hspace{0.125cm} (5) & (6)\tabnotemark{c} \\
 Index & Seyfert\,2  & \fblHe  & \ariv & \large{\ncut} & \Toiii \\ 
 & Name &  & $\frac{\lamf4711}{\lamf4740}$  &  \cmc & \degk  \\ 
\cmidrule{1-6}

1 & Mrk\,573  & 0.039 & 1.12 & $3.07\times 10^{3}$ & 13\,390 \\ 

2 & Mrk\,34  & 0.051 & 1.15 & $2.73\times 10^{3}$ & 12\,720 \\ 

3 & Mrk\,78  & 0.053 & 1.20 & $1.87\times 10^{3}$ & 12\,220 \\ 

4 & Mrk\,176  & 0.013 & 1.03 & $4.78\times 10^{3}$ & 15\,930 \\ 

5 & Mrk\,3  & 0.072 & 0.78 & $1.27\times 10^{4}$ & 14\,650 \\ 

6 & Mrk\,1  & 0.059 & 0.78 & $1.27\times 10^{4}$ & 14\,740 \\ 

7 & NGC\,1068  & 0.097 & 0.72 & $1.56\times 10^{4}$ & 14\,190 \\ 
      \bottomrule
   \end{tabular}
  \end{changemargin}
   \begin{tablenotes}
      \small
      \item $^a$Based on the reddening corrected line ratios of Table\,\ref{tab:sam}. The fits to both the \Roiii\ ratio and \ariv\ doublet assume a density distribution that extends from 100\,\cmc\ up to the cut-off density \ncut. The plasma covering factor follows a powerlaw function of density with index $\epsi\ = +0.6$.
      \item $^b$ The target \ariv\ \arivr\ ratios after the \hei\ deblending corrections have been applied to the observed values given in Col.\,(5) of Table\,\ref{tab:sam}.  
      \item $^c$ The averaged temperature for the sample is $\Tosnmn = 13\,380$\,\degk. 
   \end{tablenotes}
\end{table}

We conclude that for a covering angle $\Omega$ that increases monotonically with density, there is no evidence of the \Roiii\ ratio being affected by collisional deexcitation among the Seyfert\,2 sample of Kos78. We cannot rule out the existence of a double bump being present in the density distribution of some AGN. The cause could be the existence of a high density component above $\ga 10^6$\,\cmc\ since such component would not contribute to the \ariv\ lines and, therefore, our modelling would not be sensitive to it. There are indications that such component might be present in QSO\,2's as proposed in \S\,\ref{sec:qsotwo}. 

\end{appendices}



\setcounter{table}{2}

\begin{sidewaystable}[!t]
\caption{Exploration with \OSALD\ of blending due to \neiv\ \neivaa\   }
\label{tab:dustneon}
\begin{changemargin}{-0.5cm}{-0.5cm}
\centering
\small
    \begin{tabular}{cl ccc cccc cccc}
      \toprule
 \multicolumn{2}{c}{Objects}  & \multicolumn{3}{c}{Target line ratios}  & \multicolumn{5}{c}{Dual blending corrections\tabnotemark{a}}  & \multicolumn{3}{c}{Parameter values }\\
 \cmidrule(r){1-2} \cmidrule(lr){3-5} \cmidrule(lr){6-10} \cmidrule(lr){11-13} 

\hspace{0.015cm} (1) & \hspace{0.025cm} (2) & (3) & (4) & (5) & (6) & (7) & (8) & (9) & (10) & (11) & (12) & (13)\\
\multicolumn{1}{c}{Index} & {AGN} & \ha/\hb & \Roiii & \ariv{\tiny +}  &  \RHeAr\  & \fblHe & \RNeAr\ & \fblNe & \ariv 
& \large{\tauvi} & \large{\nopa} &  \Toiii \\

 \# &  & $\frac{\lamf4861}{\lamf6563}$ & $\frac{\lamf4363}{\lamf5007}$ & $\frac{\lamf4711{\tiny +}}{\lamf4740}$ & $\frac{\lamf5876}{\lamf4740}$ &  & $\frac{\lamf4725}{\lamf4740}$  &  & $\frac{\lamf4711}{\lamf4740}$ &  & \cmc & \degk \\ 
 \midrule

{1} & {Mrk\,573}  & {3.62} & {0.0119} & {1.156}  & {2.03} & {0.053} & {0.30\tabnotemark{b}} & {0.203} & 0.921 & {0.15} & {$3.59\times 10^{3}$} & {12\,710} \\

{2} & {Mrk\,34}  & {4.10} & {0.0110} & {1.193}  & {2.46} & {0.056} & {0.30\tabnotemark{b}} & {0.196} & 0.953 & {0.40} & {$4.56\times 10^{3}$} & {12\,610} \\

{3} & {Mrk\,78}  & {5.31} & {0.0075} & {1.238}  & {4.05} & {0.075} & {0.30\tabnotemark{b}} & {0.190} & 0.979 & {1.09} & {$6.45\times 10^{3}$} & {11\,470} \\

{4} & {Mrk\,176}  & {6.55} & {0.0139} & {1.045}  & {0.90} & {0.017} & {0.30\tabnotemark{b}} & {0.220} & 0.845 & {1.73} & {$1.47\times 10^{4}$} & {15\,120} \\
 
{5} & {Mrk\,3}  & {5.31} & {0.0141} & {0.850}  & {3.00} & {0.089} & {0.30\tabnotemark{b}} & {0.307} & 0.609 & {1.08} & {$2.76\times 10^{4}$} & {14\,390} \\

{6} & {Mrk\,1}  & {5.00} & {0.0136} & {0.814} & {2.71} & {0.089} & {0.30\tabnotemark{b}} & {0.326} & 0.575 & {0.91} & {$2.85\times 10^{4}$} & {13\,950} \\

{7} & {NGC\,1068}  & {4.47} & {0.0129} & {0.763}  & {3.56} & {0.146} & {0.30\tabnotemark{b}} & {0.373} & 0.502 & {0.61} & {$3.11\times 10^{4}$} & {13\,250}  \\




\cmidrule{1-13}

{8\hspace{0.02cm}A} & {NGC\,4151\tabnotemark{c}} & {5.29} & {0.0222} & {0.727} & {2.40} & {0.064} & {--} & {--} & 0.684 & {1.07} & {$2.16\times 10^{4}$} & {18\,050}  \\ 

{8\hspace{0.02cm}B} & \hspace{0.25cm}{\quof}\hspace{0.39cm}{\quof} & {\quof}  & {\quof} & {\quof} & {\quof} & {0.088} & {0.30\tabnotemark{d}} & {0.377} & 0.496 & {1.07} & {$4.64\times 10^{4}$} & {17\,610} \\ 

{8\hspace{0.02cm}C} & \hspace{0.25cm}{\quof}\hspace{0.39cm}{\quof} & {\quof}  & {\quof} & {\quof} & {\quof} & {0.148} & {0.62\tabnotemark{e}} & {1.32} & 0.294 & {1.08} & {$1.49\times 10^{5}$} & {16\,180} \\ 

\cmidrule{1-13}

9 & {Mrk\,477\hspace{0.03cm}} & {4.00} & {0.0215} & {0.693}  & {7.05} & {0.295} & {0.30} & {0.35} & 0.535 & {0.34} & {22\,200} & {16\,350}   \\

10 & {J1653+23\hspace{0.03cm}\tabnotemark{f}} & {4.08} & {0.0192} & {1.16}  & {4.76} & {0.099} & {0.42} & {0.25} & 1.05 & {0.39} & {2\,840} & {16\,050} \\

11 & {J1300+53\hspace{0.03cm}} & {3.79} & {0.0257} & {1.13}  & {3.21} & {0.070} & {0.38} & {0.22} & 1.06 & {0.239} & {2\,200} & {18\,320} \\

      \bottomrule
    \end{tabular}
    \begin{tablenotes}
      \small
      \item $^a$ The fraction of \arivp\ contributed by blending is the sum of $\fblHe + \fblNe$.
      \item $^b$ The quoted \neiv\ doublet ratio of 0.30 is our estimated upper limit for the Kos78 sample. 
     \item $^c$ The line ratios measurements of the Seyfert\,I NGC\,4151, are from \citet{Bok75}. 
     \item $^d$ Eye estimate of the \neiv\ \lam4725\AA\ doublet from the \citet{Bok75} spectrum. 
     \item $^e$ Value of the \neiv\ \lam4725\AA\ doublet deduced from Table\,I of  \citet{Bok75}.
     \item $^f$ Observations carried out with the spectrograph OSIRIS mounted on the 10.4\,m Gran Telescopio Canarias. 
    \end{tablenotes}
  \end{changemargin}
\end{sidewaystable}




\bibliography{NLR_binette}

\end{document}